# Classification of Major Depressive Disorder Using Vertex-Wise Brain Sulcal Depth, Curvature, and Thickness with a Deep and a Shallow Learning Model

Roberto Goya-Maldonado[1,*], Tracy Erwin-Grabner[1], Ling-Li Zeng[2,3], Christopher R. K. Ching[3], Andre Aleman[4], Alyssa R. Amod[5], Zeynep Basgoze[6], Francesco Benedetti[7], Bianca Besteher[8], Katharina Brosch[9], Robin Bülow[10], Romain Colle[11,12], Colm G. Connolly[13], Emmanuelle Corruble[11,12], Baptiste Couvy-Duchesne[14,15], Kathryn Cullen[6], Udo Dannlowski[16], Christopher G. Davey[17], Annemiek Dols[18,19], Jan Ernsting[16], Jennifer W. Evans[20], Lukas Fisch[16], Paola Fuentes-Claramonte[21], Ali Saffet Gonul[22], Ian H. Gotlib[23], Hans J. Grabe[24], Nynke A. Groenewold[5], Dominik Grotegerd[16], Tim Hahn[16], J. Paul Hamilton[25], Laura K. M. Han[26,27], Ben J. Harrison[17], Tiffany C. Ho[28,29], Neda Jahanshad[3], Alec J. Jamieson[17], Andriana Karuk[21], Tilo Kircher[9], Bonnie Klimes-Dougan[30], Sheri-Michelle Koopowitz[5], Thomas Lancaster[31,32], Ramona Leenings[16], Meng Li[8], David E. J. Linden[31,32,33,34], Frank P. MacMaster[35], David M. A. Mehler[16,31,32,36], Susanne Meinert[16,37], Elisa Melloni[7], Bryon A. Mueller[6], Benson Mwangi[38], Igor Nenadić[9], Amar Ojha[39], Yasumasa Okamoto[40], Mardien L. Oudega[41,42], Brenda W. J. H. Penninx[41], Sara Poletti[7], Edith Pomarol-Clotet[21], Maria J. Portella[43], Elena Pozzi[26,27], Joaquim Radua[44], Elena Rodríguez-Cano[21], Matthew D. Sacchet[45], Raymond Salvador[21], Anouk Schrantee[46], Kang Sim[47,48,49], Jair C. Soares[38], Aleix Solanes[44], Dan J. Stein[5], Frederike Stein[9], Aleks Stolicyn[50], Sophia I. Thomopoulos[3], Yara J. Toenders[26,27,51,52], Aslihan Uyar-Demir[22], Eduard Vieta[53], Yolanda Vives-Gilabert[54], Henry Völzke[55], Martin Walter[8,56], Heather C. Whalley[50], Sarah Whittle[17], Nils Winter[16], Katharina Wittfeld[24,57], Margaret J. Wright[56,57], Mon-Ju Wu[38], Tony T. Yang[28], Carlos Zarate[60], Dick J. Veltman[18], Lianne Schmaal[26,27], Paul M. Thompson[3], for the ENIGMA Major Depressive Disorder working group[61]

Affiliations:

[1] Laboratory of Systems Neuroscience and Imaging in Psychiatry (SNIP-Lab), Department of Psychiatry and Psychotherapy, University Medical Center Göttingen (UMG), Georg-August University, Göttingen, Germany;

[2] College of Intelligence Science and Technology, National University of Defense Technology, Changsha 410073, China;

[3] Imaging Genetics Center, Mark & Mary Stevens Neuroimaging and Informatics Institute, Keck School of Medicine, University of Southern California, Marina del Rey, CA 90274, USA;

[4] Department of Biomedical Sciences of Cells and Systems, University Medical Center Groningen, University of Groningen, Groningen, the Netherlands;

[5] Department of Psychiatry & Mental Health, Neuroscience Institute, University of Cape Town, Cape Town, South Africa;

[6] Department of Psychiatry and Behavioral Science, University of Minnesota Medical School, Minneapolis, MN, USA;

[7] Division of Neuroscience, IRCCS Scientific Institute Ospedale San Raffaele, Milano, Italy;

[8] Department of Psychiatry and Psychotherapy, Jena University Hospital, Jena, Germany;

[9] Department of Psychiatry and Psychotherapy, University of Marburg, Rudolf Bultmann Str. 8, 35039 Marburg, Germany;

[10] Institute for Radiology and Neuroradiology, University Medicine Greifswald, Greifswald, Germany;

[11] MOODS Team, CESP, INSERM U1018, Faculté de Médecine, Univ Paris-Saclay, Le Kremlin Bicêtre 94275, France;

[12] Service Hospitalo-Universitaire de Psychiatrie de Bicêtre, Hôpitaux Universitaires Paris-Saclay, Assistance Publique-Hôpitaux deParis, Hôpital de Bicêtre, Le Kremlin Bicêtre F-94275, France;

[13] Department of Biomedical Sciences, Florida State University, Tallahassee FL, USA;

[14] Sorbonne University, Paris Brain Institute - ICM, CNRS, Inria, Inserm, AP-HP, Hôpital de la Pitié Salpêtrière, F-75013, Paris, France;

[15] Institute for Molecular Bioscience, the University of Queensland, St Lucia, QLD, Australia;

[16] Institute for Translational Psychiatry, University of Münster, Münster, Germany;

[17] Melbourne Neuropsychiatry Centre, Department of Psychiatry, the University of Melbourne, Parkville, Victoria, Australia;

[18] Department of Psychiatry, Amsterdam UMC, Vrije Universiteit Amsterdam, Amsterdam Neuroscience, Amsterdam Public Health Research Institute, Amsterdam, the Netherlands;

[19] Department of Psychiatry, UMC Utrecht Brain Center, University Utrecht, Utrecht, the Netherlands;

[20] Experimental Therapeutics and Pathophysiology Branch, National Institute for Mental Health, National Institutes of Health, Bethesda, MD, USA;

[21] FIDMAG Germanes Hospitalàries Research Foundation, Centro de Investigación Biomédica en Red de Salud Mental (CIBERSAM), Instituto de Salud Carlos III, Barcelona, Catalonia, Spain;

[22] SoCAT Lab, Department of Psychiatry, School of Medicine, Ege University, Izmir, Turkey;

[23] Department of Psychology, Stanford University, Stanford, CA, USA;

[24] Department of Psychiatry and Psychotherapy, University Medicine Greifswald, Greifswald, Germany;

[25] Center for Social and Affective Neuroscience, Department of Biomedical and Clinical Sciences, Linköping University, Linköping, Sweden;

[26] Centre for Youth Mental Health, the University of Melbourne, Parkville, VIC, Australia;

[27] Orygen, Parkville, VIC, Australia;

[28] Department of Psychiatry and Behavioral Sciences, Division of Child and Adolescent Psychiatry, Weill Institute for Neurosciences, University of California, San Francisco, CA, USA;

[29] Department of Psychology, University of California, Los Angeles, CA, USA;

[30] Department of Psychology, University of Minnesota, Minneapolis, MN, USA;

[31] Cardiff University Brain Research Imaging Centre, Cardiff University, Cardiff, UK;

[32] MRC Centre for Neuropsychiatric Genetics and Genomics, Cardiff University, Cardiff, UK;

[33] Division of Psychological Medicine and Clinical Neurosciences, Cardiff University, Cardiff, UK;

[34] School of Mental Health and Neuroscience, Faculty of Health, Medicine and Life Sciences, Maastricht University, Maastricht, 6229 ER, the Netherlands;

[35] Departments of Psychiatry and Pediatrics, University of Calgary, Calgary, AB, Canada;

[36] Department of Psychiatry, Psychotherapy and Psychosomatics, Medical School, RWTH Aachen University, Germany;

[37] Institute for Translational Neuroscience, University of Münster, Germany;

[38] Center Of Excellence on Mood Disorders, Louis A. Faillace, MD, Department of Psychiatry and Behavioral Sciences at McGovern Medical School, the University of Texas Health Science Center at Houston, USA;

[39] Center for Neuroscience, University of Pittsburgh, Pittsburgh, PA, USA; Center for Neural Basis of Cognition, University of Pittsburgh, Pittsburgh, PA, USA;

[40] Department of Psychiatry and Neurosciences, Hiroshima University, Hiroshima, Japan;

[41] Department of Psychiatry, Amsterdam UMC, Vrije Universiteit Amsterdam, Amsterdam Neuroscience, Amsterdam Public Health Research Institute, Amsterdam, the Netherlands;

[42] GGZ inGeest Mental Health Care, Amsterdam, the Netherlands;

[43] Sant Pau Mental Health Research Group, Institut de Recerca de l'Hospital de la Santa Creu i Sant Pau, Barcelona, Catalonia, Spain. CIBERSAM, Madrid, Spain;

[44] Imaging of Mood- and Anxiety-Related Disorders (IMARD) Group, Institut d'Investigacions Biomèdiques August Pi i Sunyer (IDIBAPS), Barcelona, Spain.

[45] Meditation Research Program, Department of Psychiatry, Massachusetts General Hospital, Harvard Medical School, Boston, MA, USA;

[46] Amsterdam University Medical Centers, location AMC, Department of Radiology and Nuclear Medicine, Amsterdam, the Netherlands;

[47] West Region, Institute of Mental Health, Singapore;

[48] Yong Loo Lin School of Medicine, National University of Singapore, Singapore;

[49] Lee Kong Chian School of Medicine, Nanyang Technological University, Singapore;

[50] Division of Psychiatry, Centre for Clinical Brain Sciences, University of Edinburgh, Scotland, UK;

[51] Developmental and Educational Psychology, Leiden University, the Netherlands;

[52] Erasmus School of Social and Behavioral Sciences, Erasmus University Rotterdam, the Netherlands;

[53] Hospital Clinic, Institute of Neuroscience, University of Barcelona, IDIBAPS, CIBERSAM, Barcelona, Catalonia, Spain;

[54] Intelligent Data Analysis Laboratory (IDAL), Department of Electronic Engineering, Universitat de València, Valencia, Spain;

[55] Institute for Community Medicine, University Medicine Greifswald, Greifswald, Germany;

[56] Clinical Affective Neuroimaging Laboratory, Leibniz Institute for Neurobiology, Magdeburg, Germany;

[57] German Center for Neurodegenerative Diseases (DZNE), Site Rostock/ Greifswald, Greifswald, Germany;

[58] Queensland Brain Institute, the University of Queensland, Brisbane, QLD, Australia;

[59] Centre for Advanced Imaging, the University of Queensland, Brisbane, QLD, Australia;

[60] Section on the Neurobiology and Treatment of Mood Disorders, National Institute of Mental Health, Bethesda, MD, USA

[61] https://enigma.ini.usc.edu/ongoing/enigma-mdd-working-group/

***Corresponding author:**
PD Dr. Roberto Goya-Maldonado
Laboratory of Systems Neuroscience and Imaging in Psychiatry (SNIP-Lab)
Department of Psychiatry and Psychotherapy
University Medical Center Göttingen (UMG)
Von-Siebold Str. 5, 37075 Göttingen
e-mail: roberto.goya@med.uni-goettingen.de

Abstract words: 303
Text words: 5023
References: 62

Figures: 2
Tables:2

Suppl Figures: 4
Suppl Tables: 6

## Abstract

Major depressive disorder (MDD) is a complex psychiatric disorder that affects the lives of hundreds of millions of individuals around the globe. Even today, researchers debate if morphological alterations in the brain are linked to MDD, likely due to the heterogeneity of this disorder. The application of deep learning tools to neuroimaging data, capable of capturing complex non-linear patterns, has the potential to provide diagnostic and predictive biomarkers for MDD. However, previous attempts to demarcate MDD patients and healthy controls (HC) based on segmented cortical features via linear machine learning approaches have reported low accuracies. In this study, we used globally representative data from the ENIGMA-MDD working group containing 7,012 participants from 30 sites (N=2,772 MDD and N=4,240 HC), which allows a comprehensive analysis with generalizable results. Based on the hypothesis that integration of vertex-wise cortical features can improve classification performance, we evaluated the classification of a DenseNet and a Support Vector Machine (SVM), with the expectation that the former would outperform the latter. As we analyzed a multi-site sample, we additionally applied the ComBat harmonization tool to remove potential nuisance effects of site. We found that both classifiers exhibited close to chance performance (balanced accuracy DenseNet: 51%; SVM: 53%), when estimated on unseen sites. Slightly higher classification performance (balanced accuracy DenseNet: 58%; SVM: 55%) was found when the cross-validation folds contained subjects from all sites, indicating site effect. In conclusion, the integration of vertex-wise morphometric features and the use of the non-linear classifier did not lead to the differentiability between MDD and HC. Our results support the notion that MDD classification on this combination of features and classifiers is unfeasible. Future studies are needed to determine whether more sophisticated integration of information from other MRI modalities such as fMRI and DWI will lead to a higher performance in this diagnostic task.

## Introduction

Major depressive disorder (MDD) is a clinically heterogeneous psychiatric disorder manifested by low mood, anhedonia, impaired cognition, sleep disturbances, loss of energy, suicidal thoughts and appetite loss or gain. MDD dramatically impacts the daily functioning of patients and is currently the leading cause of disability worldwide (Friedrich, 2017). Therefore, early diagnosis and optimal allocation of the proper treatment are critical. Unfortunately, the current treatment strategies present a response rate and remission as low as of 36.8% after a first treatment (Machado et al., 2006; Mendlewicz, 2008; Rush et al., 2006). Thus, as proposed in the realms of systems medicine, we expect that by identifying brain patterns that classify patients at the individual level, we may open new biomarker-based avenues for the development of more personalized and effective treatments.

Neuroimaging techniques, such as magnetic resonance imaging (MRI), enable a non-invasive macro-scale view of human brain structure at the millimeter level of resolution. Initial neuroimaging studies used univariate approaches to reveal structural brain differences in MDD compared to healthy controls (HC) (Coffey et al., 1993; Sheline et al., 2003; Zhang et al., 2012), identifying reduced hippocampal and frontal lobe volume. However, these studies had limited sample sizes and the more recent large sample studies have reported small effect sizes (Ho et al., 2022; Schmaal et al., 2017, 2016; Winter et al., 2021), highlighting the absence of a single neuro-anatomical biomarker associated with MDD. The search for more complex biomarkers, which may include the interaction between different neuro-anatomical features, can be conducted via machine learning (ML) algorithms - especially deep learning (DL) algorithms - applied to the MDD vs HC classification task.

Like univariate approaches, ML and DL studies reported varying classification accuracies from 53% to 91% (Gao et al., 2018; Kambeitz et al., 2017). The high variability of classification performances and the lack of consistent biomarkers can partly be explained by the small sample sizes, as it was demonstrated by Flint and colleagues (Flint et al., 2021). Supplementing this, a study based on cortical and subcortical morphological features, reported high accuracy of 75% in the small sample, which was not replicated in an independent large UK Biobank dataset, achieving only 54% (Stolicyn et al., 2020).

Another factor that may inflate classification accuracies are related to study-site effects. The site-effect corresponds to site-specific characteristics other than diagnosis – such as scanner type, acquisition protocol, demographic differences, and inclusion and exclusion criteria – which may bias classification accuracies. A study demonstrated how site effect may contribute to both inflated and deflated classification accuracies (Solanes et al., 2021). Hence, numerous ways to tackle site-effect and improve model generalizability exist, from linear and non-linear ComBat harmonization tools (Pomponio et al., 2020; Radua et al., 2020) to embedding site confounders directly to the model (Ma et al., 2018). However, to overcome the difficult point of the heterogeneity of MDD and the lack of replicability and generalization of the models, the investigation of very large samples of participants with global representation is fundamental.
Using a large-scale dataset from the ENIGMA-MDD consortium, we compared the classification performance of commonly used ML models to predict diagnosis based on cortical and subcortical parcellations of morphological features (surface areas, thicknesses, volumes) (Belov et al., 2022). Overall, results showed a trend that may highlight the contribution of site-effects to classification performance. Specifically, there was a clear difference in classification performance dependent on the cross-validation splitting techniques used in training. Site-splitting generally performed at close to chance level for all classifiers, while mixing sites across splits achieved up to 62% balanced accuracy with an SVM. Of note, data harmonization using ComBat removed the site effect and resulted in a balanced accuracy of 52% with SVM. Based on these findings, we concluded that most commonly used ML classification algorithms could not successfully discriminate MDD from HC individuals based on morphological features organized in pre-defined Desikan-Killiany atlas parcellations. However, it remains unclear whether more fine-grained information of morphometric features, displayed in a vertex-wise organization, could outperform the classification based on parcellation atlas-distributed information.

There are some directions in improving classification based on morphological information. First, previous ML studies considered surface area, thickness, and volume characteristics only, while the information on the cortical shape , such as gyral and sulcal shape patterns, was not integrated into analyses. Cortical gyrification modalities are affected by genetic and non-genetic factors (Kremen et al., 2010; White et al., 2002), alterations of which were associated with MDD (Depping et al., 2018; Zhang et al., 2009). Multimodal morphological feature analysis, including myelination, gray matter, and curvature, revealed a correlation between cortical differences and MDD-associated genes (Li et al., 2021). Therefore, the addition of shape modalities, such as cortical curvature and sulcal depth, to cortical thickness could enhance the classification performance, as demonstrated for sex and autism classification (Gao et al., 2021).

Another direction to improve low classification performance is to deploy more advanced classification algorithms. DL methods have gained popularity in the neuroimaging field as a promising tool for cortical surface reconstruction (Cruz et al., 2021), image preprocessing (Henschel et al., 2020), and cortical parcellation (Williams et al., 2021). Furthermore, DL is widely evaluated as a predictive tool in psychiatry, showing higher or at least the same classification performance compared to linear models (Gao et al., 2021; Pinaya et al., 2018; Qin et al., 2022; Schulz et al., 2020; Wen et al., 2020; Yan et al., 2019). The analysis of cortical morphometric features can be conducted via convolutional neural network (CNN) (Lecun et al., 1998), designed to reveal complex patterns in 2D images. In order to apply such 2D CNN in the classification, it requires 3D cortical features to be initially projected into 2D image space. Nevertheless, this step may inevitably create distortion in spatial properties such as shape, area, distance, and direction. Several approaches were implemented before, such as latitude/longitude projection (Seong et al., 2018) and optimal mass transport (OMT) projection (Gao et al., 2021; Su et al., 2015), which preserves area. However, the impact of these projection methods on classification performance were never directly compared in the neuroimaging field.

The main goal of this study was to distinguish MDD from HC individuals based on integrated cortical morphological features, including sulcal depth, curvature, and thickness. These features were analyzed via SVM with linear kernel and CNN architecture of pre-trained DenseNet (Huang et al., 2017), which demonstrated its superiority over simpler models in autism vs HC classification task (Gao et al., 2021). SVM was chosen as it is a robust shallow ML model, frequently used in neuroimaging settings (Lu et al., 2016; Sacchet et al., 2015; Wottschel et al., 2019). We compared classification performance of these methods to understand the role of complex non-linear patterns in MDD manifestation. We used balanced accuracy, sensitivity, specificity and AUC as the classification performance metrics. Higher classification performance of the DenseNet model presume the presence of spatially complex patterns in brain morphology, which are relevant for classification. Furthermore, we aimed to estimate the relevance of integrating cortical thickness and shape characteristics (sulcal depth, curvature and thickness) into the analysis by training the models with all features combined and by considering them separately. Similar to our previous study (Belov et al., 2022), different cross-validation (CV) approaches were evaluated: Splitting the data by balancing age and sex distribution across all CV folds (Splitting by Age/Sex), and performing leave-sites-out CV in order to estimate the performance on the unseen during the training sites (Splitting by Site). This approach allowed us to estimate whether the model's performance is influenced by demographic or site-related factors. The difference between results in both splitting strategies presumes the presence of the site effect we addressed by harmonizing the data in both splitting strategies via ComBat. In summary, we hypothesized that: (1) Integration of cortical thickness and shape characteristics would contribute positively to the classification performance, and (2) DenseNet could differentiate MDD from HC based on the provided features. Additionally, we compared two projection methods, latitude/longitude and OMT projections by performing auxiliary single-site sex classification based on three of the largest cohorts to explore whether classification performance may vary according to 2D projection method. We had no *a priori* hypothesis for the projection results.

## Material and methods

*Study participants and study design*

We analyzed a large-scale multi-site sample provided by ENIGMA-MDD working group, comprising 2,772 MDD and 4,240 HC individuals, from 30 cohorts worldwide. Details on inclusion/exclusion criteria and sample characteristics can be found in **Supplementary Table 1**. Subjects with missing information on demographic data or any of cortical surface mesh files (l(r).sulc, l(r).curv, l(r).thickness) were excluded from the analysis (476 and 6 % excluded). The study was approved by the Ethics Committee of the University Medical Center (UMG), Germany. In accordance with the Declaration of Helsinki, all participating cohorts confirmed approval from their corresponding institutional review boards and local ethics committees as well as collected written consent of all participants. In case of participants under 18 years old, the written consent was also given by a parent and/or legal guardian.

*Image processing and analysis*

Each site acquired structural T1-weighted MRI scans of participants and preprocessed them according to ENIGMA Consortium protocol (http://enigma.ini.usc.edu/protocols/imaging-protocols/). This pipeline includes the segmentation of T1-weighted MRI volumes, tessellation, topology correction, and spherical inflation of the white matter surface. Detailed information on the acquisition protocols and scanner model in each cohort can be found in **Supplementary Table 2**. Cortical meshes were generated during FreeSurfer preprocessing in every site. Cerebral cortex meshes were then extracted from the FreeSurfer unsmoothed fsaverage6 template, effectively removing intracranial volume (ICV) differences (**Supplementary Figure 1**) and yielding 37,747 and 37,766 vertices for the left and right hemispheres, respectively. We analyzed vertex-wise features, such as sulcal depth, curvature, and thickness, both as integrated features and separately (**Figure 1**).

Considering the absence of well-established pre-trained on cortical meshes CNN models, we projected 3D cortical surfaces into 2D images and applied DenseNet, which was pre-trained on natural images. There are few studies applying different projection methods such as latitude/longitude project and area-preserving maps (e.g., Seong et al., 2018; Gao et al., 2021). Of note, the latitude/longitude method, in which cortical mesh is first re-sampled to the sphere and consequently mapped to the 2D grid, creates strong area distortions in the edges and near the medial wall close to subcortical regions (Seong et al., 2018). Both methods may (differentially) influence subsequent classification performances, but to the best of our knowledge, no studies to date have directly compared this in one study using the same samples. Thus, we applied both 2D projection methods to the cortical meshes, resulting in 224×224 pixels images for each hemisphere. The images were normalized to present mean of 0 and standard deviation of 1.

*Data Splitting*

To assess potential biases in the model's decision-making, we performed 10-fold cross-validation (CV) by splitting the data according to 1) demographic covariates, in which age and sex distribution were balanced and subjects from each site are equally distributed across all CV folds (*Splitting by Age/Sex*), and 2) site affiliation, where each site was contained only in one CV fold (*Splitting by Site*). In both strategies, 9 CV folds were used for training, while one remaining CV fold was used as a test set. This procedure was repeated iteratively until every CV fold was used as a test set. In the Splitting by Age/Sex strategy, effect of demographic factors on the classification performance is reduced, as the model is trained and tested on the same demographics. Nevertheless, the site-related differences may bias the decision-making of the classification models (Belov et al., 2022), which is directly addressed in Splitting by Site. This strategy demonstrates how well the model trained on one set of sites can be applied to the data from unseen sites. As the number of sites exceeds the number of folds, we distributed the sites across the folds to balance the number of subjects in every fold as close as possible by iteratively distributing the largest sites across all 10 folds. Smallest folds were added subsequently to further even the number of subjects in every fold. Overall, the difference in the classification results between these two splitting strategies may indicate the existence of the site effect. More detailed description of both splitting strategies can be found elsewhere (Belov et al., 2022).

*MDD vs HC classification*

After the data-splitting step, the primary analysis was carried out. Firstly, we residualized all features normatively, removing linear age and sex dependencies. To avoid data leakage, age and sex regressors were estimated on the healthy subjects from the training set (9 CV folds) and then applied to the training and test set (1 CV fold) for patients and HC. Next, the classification algorithms were trained on the training folds, and classification performance was estimated on the test fold. As demonstrated by Dinga and colleagues, accuracy alone should be avoided as it does not account for class frequencies (Dinga et al., 2019). Thus, the algorithms were evaluated according to categorical measures, including balanced accuracy, sensitivity, specificity, and rank-based measure – AUC, allowing for a broad overview of performance. For model-level assessment (Kohoutová et al., 2020), we performed the classification using all features combined and then using features separately to assess the final classification performance. We evaluated the classification performance of a robust shallow model - SVM with linear kernel, and deep learning model - DenseNet pre-trained on natural images from ImageNet dataset (Deng et al., 2009), which has been demonstrated as a robust convolutional neural network for image classification both for natural images as well as in neuroimaging (Gao et al., 2021; Huang et al., 2017). When DenseNet is trained on a single data domain, left and right hemisphere images are propagated through corresponding left and right DenseNets, the fully connected layers of which are concatenated. The resulting feature vectors are fed to the output layer. For the whole-brain all-features analysis, we combined the features extracted from every feature and hemisphere, concatenate them and feed them to the output layer. For SVM, all considered images were flattened and then concatenated into a single array. The concept and implementation of analysis is presented in **Figure 1**. To mitigate site-related differences, which may potentially bias the classification results, we additionally performed the analysis with harmonizing all of the features via ComBat. Variance explained by age and sex was preserved during this harmonization step. Next, we residualized features normatively, as described above, and train/test the models. Application of ComBat differed for both splitting strategies. In short, ComBat parameters estimated on the training set were applied to the test set directly, in the splitting by Age/Sex. In splitting by Site, ComBat is applied twice. Firstly, we use ComBat to harmonize the training sites. Secondly, we apply ComBat to adjust test sites to the harmonized training sites, i.e. using the training sites as the reference batch (Zhang et al., 2018). A more detailed description of ComBat application can be found in our previous work (Belov et al., 2022).

*Auxiliary analysis in projection methods*

To explore and evaluate the potential impact of 2D projection methods on the classification performance, we compared both methods in their ability to classify healthy males from healthy females in 3 of the largest cohorts separately. The single-site classification was estimated via 10-fold CV on 411, 723, and 397 subjects, respectively. As usual, 9 CV folds were used for training, while one remaining CV fold was used as a test set. This procedure was repeated iteratively until every CV fold is used as a test set. In order to obtain an initial view of pre-trained DenseNet, we evidenced the balanced accuracies of two models: a SVM with linear kernel and a pre-trained DenseNet (Huang et al., 2017). Furthermore, we used sex classification task to find the optimal hyperparameters for both SVM and DenseNet (**Supplementary Table 3**). Finally, to examine the possible advantage of using SVM and the pre-trained DenseNet in the sex classification task, we compared the classification performance of both models.

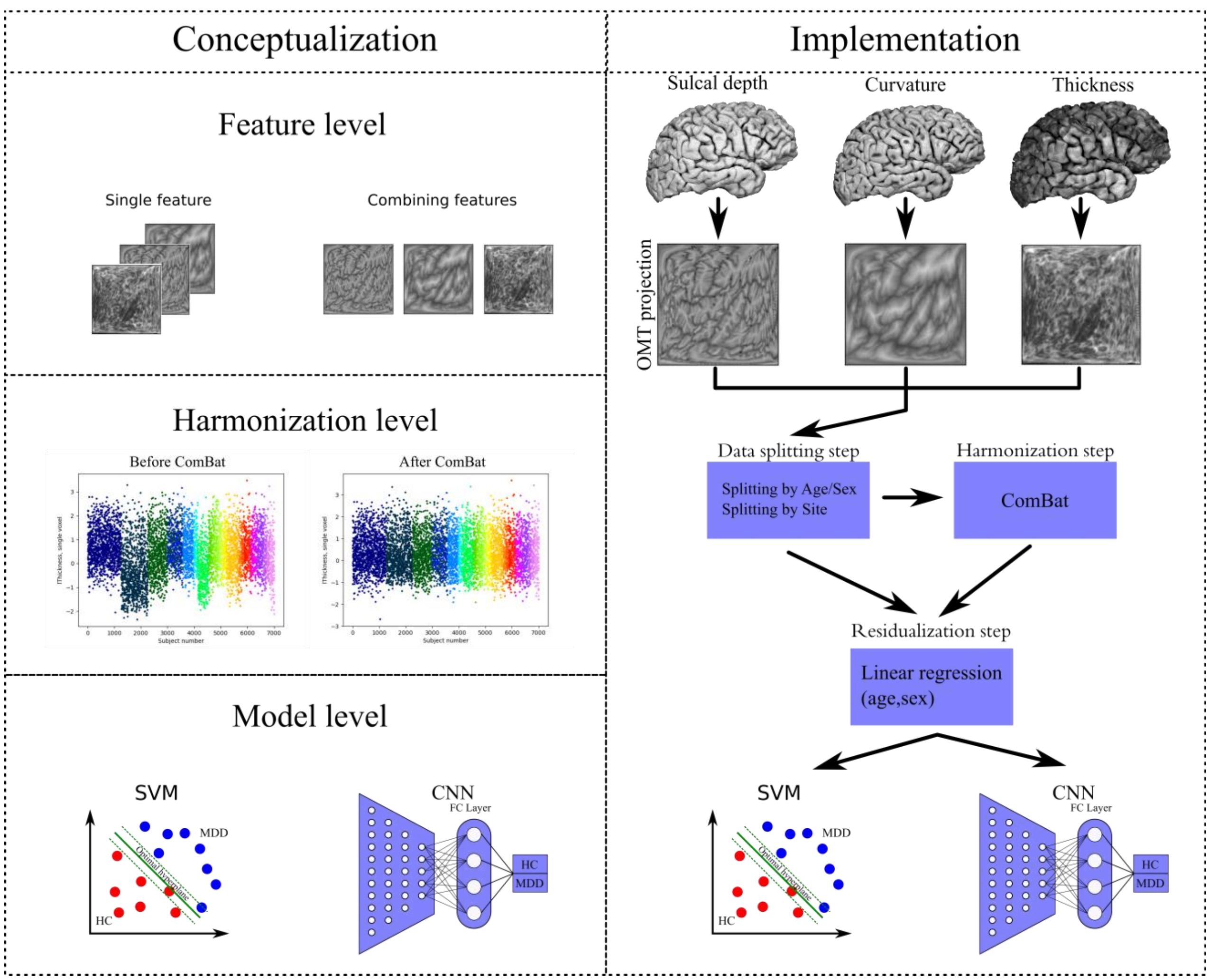

**Figure 1: Proposed conceptualization levels and implementation of classification procedure. Left:** Higher classification performance in MDD vs HC classification task can be achieved by implementing deep ML models, such as DenseNet, in comparison to a shallow ML model, for example, SVM. Furthermore, the analysis of integrated morphometric features can provide a more detailed description of cortical organization than separated features, leading to better differentiability of MDD from HC. The application of ComBat may improve the generalizability of results as site-related differences are removed. **Right:** Cortical sulcal depth, curvature, and thickness are first projected into the 2D grid and then transformed into 2D images using OMT projection. We split the data into 10 CV folds according to age and sex (Splitting by Age/Sex) and according to the site belonging (Splitting by Site). After the residualization step, where the age and sex effect are regressed out linearly, we train and test SVM and DenseNet on the diagnosis classification.

## Results

*Participants and Data Splitting*

We detected substantial differences in age (78% of pairwise comparisons between cohorts were significant, t-test, $p<0.05$) and sex (47%, t-test, $p<0.05$) across cohorts. The full demographic profile is presented in **Table 1**. As expected, Splitting by Age/Sex resulted in more balanced CV folds with respect to number of subjects, age and sex distributions, while folds created by Splitting by Site were more uneven on these characteristics (**Table 2**).

**Table 1: Participating sites.** The total number of subjects, number of MDD patients and number of HCs, as well as their mean age (in years) and sex (number and % of females) is presented.

| Cohort | | Number of subjects | Age Mean (SD) | Number of Females (%) |
|---|---|---|---|---|
| **AFFDIS** | **Total** | 79 | 39.75 (14.67) | 36(46) |
| | **HC** | 46 | 39.87 (14.29) | 22(48) |
| | **MDD** | 33 | 39.58 (15.18) | 14(42) |
| **Barcelona-StPau** | **Total** | 94 | 46.66 (7.81) | 72(77) |
| | **HC** | 32 | 46.03 (8.00) | 23(72) |
| | **MDD** | 62 | 46.98 (7.68) | 49(79) |
| **CARDIFF** | **Total** | 40 | 46.55 (11.74) | 27(68) |
| | **HC** | 0 | nan | nan |
| | **MDD** | 40 | 46.55 (11.74) | 27(68) |
| **CSAN** | **Total** | 109 | 34.70 (12.88) | 74(68) |
| | **HC** | 49 | 33.20 (12.07) | 34(69) |
| | **MDD** | 60 | 35.92 (13.38) | 40(67) |
| **Calgary** | **Total** | 107 | 17.03 (4.12) | 60(56) |
| | **HC** | 52 | 15.81 (5.03) | 29(56) |
| | **MDD** | 55 | 18.19 (2.51) | 31(56) |
| **DCHS** | **Total** | 79 | 30.91 (6.71) | 79(100) |
| | **HC** | 61 | 31.49 (6.82) | 61(100) |
| | **MDD** | 18 | 28.94 (5.89) | 18(100) |
| **FIDMAG** | **Total** | 69 | 47.22 (12.29) | 44(64) |
| | **HC** | 34 | 45.94 (11.49) | 22(65) |
| | **MDD** | 35 | 48.46 (12.90) | 22(63) |
| **FOR2107Marburg** | **Total** | 738 | 36.30(13.39) | 465 (63) |
| | **HC** | 411 | 34.76(12.76) | 257(63) |
| | **MDD** | 327 | 38.24(13.91) | 208(63) |
| **FOR2107Munster** | **Total** | 395 | 31.66(12.09) | 249 (63) |
| | **HC** | 221 | 28.39(10.29) | 140 (63) |
| | **MDD** | 174 | 35.87(12.84) | 109(63) |
| **Houston** | **Total** | 290 | 28.72 (16.30) | 169(58) |
| | **HC** | 186 | 26.76 (15.91) | 105(56) |
| | **MDD** | 104 | 32.23 (16.39) | 64(62) |
| **Hiroshima** | **Total** | 319 | 41.93(12.36) | 175(55) |
| | **HC** | 169 | 39.87(12.36) | 104(62) |
| | **MDD** | 150 | 44.24(11.94) | 71(47) |
| **Jena** | **Total** | 107 | 46.76 (15.00) | 52(49) |
| | **HC** | 77 | 47.75 (15.93) | 36(47) |
| | **MDD** | 30 | 44.20 (11.92) | 16(53) |
| **MODECT** | **Total** | 42 | 72.71 (9.25) | 28(67) |
| | **HC** | 0 | nan | nan |
| | **MDD** | 42 | 72.71 (9.25) | 28(67) |
| **Melbourne** | **Total** | 245 | 19.42 (2.88) | 130(53) |
| | **HC** | 102 | 19.58 (2.97) | 54(53) |
| | **MDD** | 143 | 13.31 (2.80) | 76(53) |
| **Minnesota** | **Total** | 110 | 15.47 (1.89) | 79(72) |
| | **HC** | 40 | 15.68 (1.98) | 26(65) |
| | **MDD** | 70 | 15.36 (1.83) | 53(76) |
| **MOODS** | **Total** | 96 | 34.54(12.48) | 65(68) |
| | **HC** | 32 | 38.87(12.36) | 104(62) |
| | **MDD** | 64 | 44.25(11.95) | 71(47) |
| **Moraldilemma** | **Total** | 70 | 18.81 (1.94) | 70(100) |
| | **HC** | 46 | 18.50 (1.75) | 46(100) |
| | **MDD** | 24 | 19.42 (2.14) | 24(100) |
| **NESDA** | **Total** | 219 | 38.11 (10.32) | 145(66) |
| | **HC** | 65 | 40.29 (9.67) | 42(65) |
| | **MDD** | 154 | 37.19 (10.45) | 103(67) |
| **QTIM** | **Total** | 386 | 22.08 (3.25) | 267(69) |
| | **HC** | 284 | 22.11 (3.30) | 190(67) |
| | **MDD** | 102 | 22.01 (3.11) | 77(75) |
| **UCSF** | **Total** | 163 | 15.46 (1.31) | 91(56) |
| | **HC** | 88 | 15.32 (1.28) | 42(48) |
| | **MDD** | 75 | 15.63 (1.33) | 49(65) |
| **SHIP_START_2** | **Total** | 579 | 55.01 (12.57) | 294(51) |
| | **HC** | 443 | 55.44 (12.80) | 198(45) |
| | **MDD** | 136 | 53.59 (11.68) | 96(71) |
| **SHIP_TREND_0** | **Total** | 1229 | 50.15 (13.69) | 607(49) |
| | **HC** | 919 | 50.50 (14.18) | 405(44) |
| | **MDD** | 310 | 49.12 (12.04) | 202 (65) |
| **SanRaffaele** | **Total** | 45 | 49.07 (13.51) | 32(71) |
| | **HC** | 0 | nan | nan |
| | **MDD** | 45 | 49.07 (13.51) | 32(71) |
| **Sexpect** | **Total** | 40 | 36(9.69) | 11 (27) |
| | **HC** | 20 | 33.75(7.02) | 3(15) |
| | **MDD** | 20 | 38.25(11.34) | 8(40) |
| **Singapore** | **Total** | 38 | 39.50 (6.43) | 18(47) |
| | **HC** | 16 | 38.69 (4.59) | 8(50) |
| | **MDD** | 22 | 40.09 (7.43) | 10(45) |
| **Socat_dep** | **Total** | 179 | 37.85 (13.34) | 161(90) |
| | **HC** | 100 | 36.42 (13.57) | 90 (90) |
| | **MDD** | 79 | 39.66 (12.81) | 71 (90) |
| **StanfFAA** | **Total** | 32 | 32.71 (9.56) | 32(100) |
| | **HC** | 18 | 30.44 (9.96) | 18(100) |
| | **MDD** | 14 | 35.63 (8.14) | 14(100) |
| **StanfT1wAggr** | **Total** | 115 | 37.18 (10.27) | 69(60) |
| | **HC** | 59 | 37.24 (10.43) | 36(61) |
| | **MDD** | 56 | 37.11 (10.09) | 33(59) |
| **TAD** | **Total** | 39 | 16.03(1.14) | 11(27) |
| | **HC** | 0 | nan | nan |
| | **MDD** | 39 | 16.03(1.14) | 11(27) |
| **TIGER** | **Total** | 60 | 15.63 (1.34) | 38(63) |
| | **HC** | 11 | 15.18 (1.03) | 5(45) |
| | **MDD** | 49 | 15.73 (1.38) | 33(67) |
| **All sites** | **Total** | 7012 | 38.41(16.28) | 4186(60) |
| | **HC** | 4240 | 39.98(14.46) | 2383(59) |
| | **MDD** | 2772 | 39.57(15.28) | 1803(61) |

**Table 2: Data splitting strategies.** Differences manifested in age/sex distribution and number of subjects between corresponding folds per splitting strategy.

| **Splitting By Age/Sex** | | | | **Splitting by Site** | | | |
|---|---|---|---|---|---|---|---|
| **Fold** | **Number of subjects** | **Mean age (SD)** | **Number of Females (%)** | **Fold** | **Number of subjects** | **Mean age (SD)** | **Number of Females (%)** |
| **0** | 708 | 38.34 (16.41) | 434 (61) | **0** | 1249 | 50.28 (13.78) | 612 (49) |
| **1** | 685 | 38.41 (16.51) | 395 (58) | **1** | 1005 | 36.01 (12.14) | 577 (57) |
| **2** | 692 | 38.59 (16.25) | 441 (64) | **2** | 738 | 36.30 (13.39) | 465 (63) |
| **3** | 709 | 37.99 (16.07) | 428 (60) | **3** | 579 | 55.00 (12.57) | 294 (51) |
| **4** | 704 | 38.74 (15.93) | 417 (59) | **4** | 563 | 33.06 (15.73) | 374 (66) |

| 5 | 708 | 38.90 (16.28) | 415 (58) | 5 | 596 | 26.42 (11.25) | 370 (62) |
|---|---|---|---|---|---|---|---|
| 6 | 693 | 38.09 (16.27) | 423 (61) | 6 | 559 | 36.89 (13.71) | 372 (67) |
| 7 | 716 | 38.3 (16.35) | 431 (60) | 7 | 589 | 35.71 (16.52) | 356 (60) |
| 8 | 689 | 38.55 (16.12) | 396 (57) | 8 | 546 | 28.70 (13.59) | 359 (66) |
| 9 | 708 | 38.14 (16.57) | 406 (57) | 9 | 588 | 33.99 (16.12) | 407 (69) |

*MDD vs HC classification*

First, we compared the performance of SVM and DenseNet for different splitting strategies (**Figure 2**). In Splitting by Age/Sex, SVM achieved 0.551±0.021 in balanced accuracy, while DenseNet yielded 0.578 ± 0.022. In Splitting by Site, both SVM and DenseNet models performed worse, yielding 0.528 ± 0.039 and 0.512 ± 0.019, respectively. The minor difference in classification performances for different splitting strategies indicated a potential site effect, which we addressed by applying ComBat. In Splitting by Age/Sex, the balanced accuracy of SVM with ComBat dropped to 0.478 ± 0.019, while the performance of DenseNet did not change and yielded 0.561 ± 0.015. In splitting by Site with ComBat, the performance of both models was similar and close to random chance, balanced accuracy yielded 0.520 ± 0.019 and 0.508 ± 0.020 for SVM and DenseNet respectively. Thus, we did not observe an improvement of models performances after data harmonization by ComBat. A full panel of results, including all classification metrics, can be found in **Supplementary Table 4**.

Next, we explored if any of the considered feature modalities yields greater classification performance (**Figure 2**). In Splitting by Age/Sex, all data modalities yielded similar range of accuracies: thickness (SVM: 0.549 ± 0.020; DenseNet: 0.576 ± 0.019) compared to sulcal depth (SVM: 0.543 ± 0.022; DenseNet: 0.562 ± 0.019), and curvature (SVM: 0.531 ± 0.015; DenseNet: 0.567± 0.019), observed for both classification models. In Splitting by Site, sulcal depth (SVM: 0.523 ± 0.016; DenseNet: 0.515 ± 0.020), curvature (SVM: 0.513 ± 0.033; DenseNet: 0.516 ± 0.025) and thickness (SVM: 0.522 ± 0.038; DenseNet: 0.515 ± 0.022) also exhibited similar range of classification accuracies. Both models performed similarly for all feature types. These results demonstrate that integration of shape modalities with cortical thickness did not benefit the classification models. Results from explorative analyses for each hemisphere and for each feature modality per hemisphere showed no improvements in performance of the models (**Supplementary Table 5**, **Supplementary Figure 3**). In addition, we applied the main demographic and clinical stratifications used in the ENIGMA-MDD working group to assess post-hoc whether groups that are more homogeneous would achieve better classification metrics (**Supplementary Table 6**).

*Auxiliary sex prediction task*

As an initial step, we also conducted a sex classification to explore, which projection method (latitude/longitude, OMT) yields higher classification performance for both SVM and DenseNet (**Supplementary Figure 2**). There was no clear difference between projection methods; however, we observed a consistently higher classification performance of DenseNet compared to SVM for all types of features and hemispheres. Considering previous success of OMT projection as a projection method applied on cortical surface and its property to preserve distances between vertices (Gao et al., 2021), we conducted our main analysis with OMT projection.

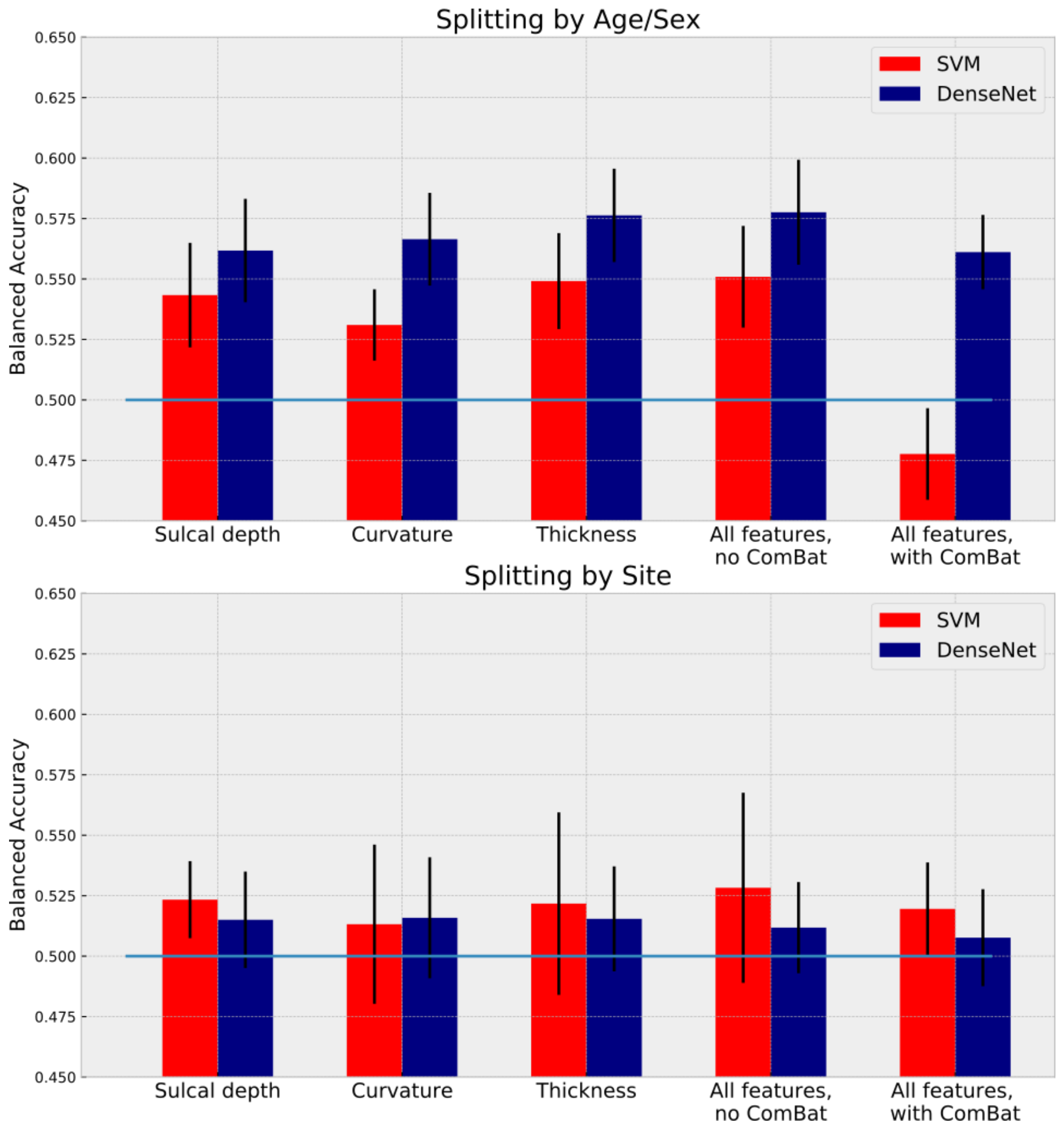

**Figure 2: MDD vs HC classification performance of SVM and DenseNet applied to vertex-wise cortical features.** Balanced accuracy for both classification models when trained on all features integrated with and without ComBat harmonization for both splitting strategies and when trained on single features. Error bars represent standard deviation.

# Discussion

In this work, we evaluated the diagnostic classification performance of DenseNet and SVM models, trained on cortical maps projected via OMT, including sulcal depth, curvature, and thickness, from a multi-site global dataset. Our analysis included 7,012 participants from 30 sites worldwide, allowing for a comprehensive and realistic overview of classification performances. Both models were evaluated in parallel using two different CV splitting strategies. In Splitting by Age/Sex, we obtained CV folds with comparable demographics; thus, the performance of the models should not be affected by these demographic variables. In Splitting by Site, sites were distributed across folds. Therefore, models were trained and tested on different sets of sites. This strategy is closer to application of diagnostic classification models in clinical practice, and allowed for realistic estimation of classification performance on unseen sites. Overall, the classification performances of both models were similar: In Splitting by Age/Sex, DenseNet achieved 58% vs 55% for SVM; in Splitting by Site, the difference was even more negligible, DenseNet achieved 51% vs 52% for SVM. Both models performed better in Splitting by Age/Sex, implying the presence of a confounding site effect, most likely arising from differences in scanner vendors or image acquisition parameters. In this sense, ComBat approximated the classification results of the two splitting strategies, but did not improve the accuracy of the models. Ultimately, the classification performances of both models for all integrated morphometric features, both in Splitting by Age/Sex and in Splitting by Site, revealed similar classification levels of single-features.

*Cortical morphological maps as diagnostic biomarkers for MDD*

To the best of our knowledge, this the first study to combine cortical thickness, sulcal depth, and curvature features in order to classify MDD vs HC. Furthermore, previous ML studies with large samples only incorporated low-resolution atlas-based thickness characteristics. In our approach, we analyzed vertex-wise information, providing a richer and more detailed description of brain characteristics than atlas-derived regional measures. Even so, the integration of complementary cortical characteristics did not lead to higher classification performances compared to the accuracies obtained from the single cortical features, regardless of the data splitting strategy and the classification model. In Splitting by Site, no feature yielded an accuracy substantially higher than random chance accuracy, indicating the failure of both models to capture MDD-specific alterations. Furthermore, the analysis of finer-grained cortical maps, even for thickness alone, did not result in higher classification performance, compared to ML performance levels observed in our previous study (Belov et al., 2022). Thus, the assumption that higher resolution would lead to greater classification performance did not hold in this study, as all results were close to the chance level, in line with previous attempts in classifying MDD (Belov et al., 2022; Flint et al., 2021; Stolicyn et al., 2020). Furthermore, stratification of the sample according to demographic (sex) and clinical characteristics (age of onset, antidepressant use, and number of depressive episodes) did not yield better differentiability between HC and MDD, in line with our previous study (Belov et al., 2022). This new evidence suggests the absence of prominent gray matter alterations that alone may serve as diagnostic tool in patients with MDD.

Although we combined complementary characteristics in the analysis, the interaction between thickness and shape was not addressed here. According to recent evidence, local cortical shape may correlate with thickness (Demirci and Holland, 2022). So, combined thickness-shape patterns should be further explored for the classification of MDD. Furthermore, reduced myelination was associated with MDD (Ho et al., 2021; Sacchet and Gotlib, 2017; van Velzen et al., 2020), which could lead to structural reorganization of cortical features, making it a potential classification aspect to be investigated. In addition, subcortical morphological characteristics may improve the classification by taking into account structural modifications in cortico-subcortical loops associated with MDD (Ho et al., 2022).

Integration of morphological characteristics with cytoarchitectonic and functional information may allow better contextualization of MDD-related alterations, as demonstrated in transdiagnostic study (Hettwer et al., 2022), with the potential to achieve higher classification performance (Ayyash et al., 2021; Zhang et al., 2020). Brain topology can be described via the connectome - a whole-brain connectivity architecture of the brain. As nodes of brain connectome exhibited elevated susceptibility to brain disorders (Fornito et al., 2015), graph analytical approaches could also lead to stronger differentiability between MDD and HC. Moreover, subject-specific parcellation schemes could be applied to compute structural and functional connectomes (Wig et al., 2013), and further analyzed by suitable sophisticated classification models taking into account the neural architecture e.g., with graph neural network (Zhou et al., 2019).

*Data Splitting and Site Effect*

Several multi-site psychiatric neuroimaging studies directly demonstrated how different splitting strategies might introduce unwanted biases in inflated classification performances (Belov et al., 2022; Mårtensson et al., 2020; Qin et al., 2022). In Splitting by Age/Sex, trained models are unbiased regarding demographic factors; while in Splitting by Site the site affiliation is controlled, therefore we addressed the generalizability of the models applied to unseen sites. Similar to the results from our previous study (Belov et al., 2022), the classification performance of both SVM and DenseNet was higher in Splitting by Age/Sex, up to 58%, compared to Splitting by Site, close to random chance. This discrepancy indicates the existence of hidden site-related biases influencing classification performance. As this nuisance-based phenomenon appears in multi-site mega-analyses (Nakano et al., 2020; Qin et al., 2022), for its better comprehension, we strongly encourage the application of different splitting strategies in future multi-site ML studies.

The low accuracy of both models in Splitting by Site strategy is either due to the presence of a strong site-effect, hindering the ability of the models to capture diagnosis-related differences, or due to the general inability of both models to find meaningful alterations associated with MDD. Therefore, we addressed site-effect via ComBat. Thus, the possibility remains that subject-level prediction based on cortical features is unfeasible. As Combat has never been applied to vertex-wise cortical projections, we visually inspected its effect on a single pixel for every feature type (**Supplementary Figure 4**). The application of ComBat resulted in more homogenous value distribution across cohorts, in line with previous studies analyzing the effects on atlas-based features (Belov et al., 2022; Pomponio et al., 2020). Nevertheless, this harmonization step did not lead to improvement in accuracies. While demographic covariates were preserved, ComBat may over-correct the data (Bayer et al., 2022), causing a part of MDD-related associations to be removed along with the site-effect. Against this, more careful consideration of the site-effect is required in the future studies.

In Splitting by Age/Sex, the balanced accuracy of both models dropped (SVM: 55% to 48%; DenseNet: 58% to 56%) when ComBat was applied. The decrease of model's performances near the levels in Splitting by Site indicates that initial higher classifications are most likely driven by site-related biases. To further validate this assumption, we performed the classification with balanced ratio between HC and MDD in every site in Splitting by Age/Sex, which resulted in close to random chance accuracies in DenseNet and SVM. Noticeably, DenseNet was less affected by the application of ComBat in the original analysis, reflecting potential non-linear site-related differences that remained in the dataset after harmonization, which is in line with previous findings (Solanes et al., 2022). Therefore, we recommend

ComBat only be applied when combining more linear models, such as SVM, while more sophisticated models alone should directly incorporate site information as an additional input.

*SVM vs DenseNet*

Previous ML mega-analyses based on structural MDD vs HC classifications considered only shallow linear and non-linear ML models, such as SVM, penalized logistic regression and decision tree (Belov et al., 2022; Flint et al., 2021; Stolicyn et al., 2020). In this study, we extended the diagnostic classification approach by comparing the performance of shallow linear model - SVM with a linear kernel to a highly non-linear deep DenseNet classifier applied to vertex-wise cortical information. The explorative results of sex classification applied to HC revealed higher classification performance of the DenseNet compared to the SVM (**Supplementary Figure 2**) for all data modalities. The higher accuracy suggests that DenseNet was able to capture non-linear sex dependencies that were present in the cortical maps. The superiority of DenseNet over SVM in the sex classification task was in line with previous study conducted on the same vertex-wise cortical maps (Gao et al., 2021). Conversely, another large sample study revealed no advantage of using any deep architectures over simpler models in predicting demographic factors (Schulz et al., 2020); therefore further tests in even bigger samples are required. Nevertheless, in this study both models exhibited a similar range of accuracies, close to random chance, for the main task of MDD versus HC classification. Therefore, the application of DenseNet did not yield the expected improvement for detecting combined (nor separated) structural cortical features that discriminate patients from controls.

Similar performance of the linear SVM and non-linear DenseNet model may be due to the absence of non-linear interactions between different cortical regions, significant for the MDD detection. Furthermore, the analyzed sample is highly heterogeneous in terms of demographic and clinical covariates, potentially interfering with the main task and lowering the classification performance. In this vein, there are several possible directions for improving DenseNet performance. First, the considered model was pre-trained only on natural images from ImageNet. The model could be subsequently pre-trained on cortical projections from an independent large sample using immediate task, for example predicting sex as it was performed in Gao's study (Gao et al., 2021). Furthermore, one could use more than one intermediate task to optimize the weights of the neural network, for example, predicting demographic or clinical covariates. This approach is broadly known as multi-task learning (Caruana, 1997), the usefulness of which in the neuroimaging domain was already demonstrated (Ma et al., 2018; Pinaya et al., 2018).

Secondly, the multi-task approach could be used to “unlearn” undesired biases. In our analysis, site-related differences were removed via ComBat. One could train the network to perform the main task while unlearning the scanner parameters, as was successfully demonstrated by Dinsdale and colleagues (Dinsdale et al., 2020). Furthermore, one could replace the residualization step in the same manner by making the network unlearn age and sex dependencies. In line with our previous analysis, we linearly regressed out age and sex dependencies from the cortical features using normative approach (Belov et al., 2022). Considering the greater performance of the DenseNet model in predicting sex, we can speculate the presence of non-linear male-female differences in cortical morphology. Thus, unlearning age- and sex-related dependencies could improve classification performance.

## Conclusion

In this study, we tested if more advanced classification algorithms applied to high-resolution morphometric shape characteristics can improve MDD vs HC classification. Splitting the data according to demographic variables and according to site allowed a comprehensive analysis of model’s performances and biases. We detected site effects, which we addressed at least partially with the ComBat harmonization tool, but did not increase classification metrics. Both shallow and deep ML models exhibited low, close to chance accuracies. Most importantly, the integration of high-resolution cortical thickness and shape features from vertices did not lead to greater classification performance over previously analyzed atlas-based cortical features. According to our results, it seems unlikely that structural MRI alone will provide diagnostic biomarkers of MDD. Thus, further investigation is required into the classification performance applied to the fusion of other MRI modalities, including fMRI and DWI.

## Declaration of Competing Interest

PMT and NJ received a research grant from Biogen, Inc. for research unrelated to this manuscript. HJG has received travel grants and speakers’ honoraria from Fresenius Medical Care, Neuraxpharm, Servier and Janssen Cilag as well as research funding from Fresenius Medical Care unrelated to this manuscript. JCS has served as a consultant for Pfizer, Sunovion,

Sanofi, Johnson & Johnson, Livanova, and Boehringer Ingelheim. The remaining authors declare no conflict of interest.

## Data availability

The datasets generated and/or analyzed during the current study are not publicly available due to site restrictions but data may be available from the corresponding sites on reasonable request.

## Contributions

All authors listed here meet the criteria for authorship, having made substantial contributions to the conception and design of the study, the acquisition of data, or the analysis and interpretation of the data; participated in drafting the manuscript or critically revising it for significant intellectual content; and approved the final version to be published.

## Acknowledgements

We would like to sincerely thank Vladimir Belov for his invaluable contributions to this work. ENIGMA MDD: This work was supported by NIH grants U54 EB020403 (PMT), R01MH116147 (PMT), and R01 MH117601 (NJ & LS), and the NSFC grants 61722313 (LLZ), and 62036013 (LLZ). LH was funded by the Rubicon award (grant number 452020227) from the Dutch NOW. AFFDIS: This study was funded by the University Medical Center Göttingen (UMG Startförderung) and VB and RGM are supported by German Federal Ministry of Education and Research (Bundesministerium für Bildung und Forschung, BMBF: 01 ZX 1507, "PreNeSt - e:Med"). Calgary: This study was supported by Alberta Children's Hospital Foundation, Canadian Institutes for Health Research. Cardiff: This work was supported by a Medical Research Council (G 1100629) grant to DEJ Linden and a PhD studentship by Health Research Wales (HS/14/20) for DMEM. CSAN: This work was supported by grants from Johnson & Johnson Innovation (S.E.), the Swedish Medical Research Council (S.E.: 2017–00875, M.H.: 2013–07434, 2019–01138), the ALF Grants, Region Östergötland M.H., J.P.H.), National Institutes of Health (R.D.: R01 CA193522 and R01 NS073939), MD Anderson Cancer Support Grant (R.D.: P30CA016672). Dep-arrest clin: BCD is supported by a NHMRC CJ Martin fellowship (app 1161356). DCHS: supported by the Medical Research Council of South Africa. The DCHS was funded by the Bill and Melinda Gates Foundation (OPP1017641), and received additional support from the South African Medical Research Council. ENIGMA Central/USC: This work was supported by the National Natural Science Foundation of China (61722313 and 62036013). ETPB: Funding for this work was provided by the Intramural

Research Program at the National Institute of Mental Health, National Institutes of Health (IRP-NIMH-NIH; ZIA-MH002857), by a NARSAD Independent Investigator to Dr. Zarate, and by a Brain & Behavior Mood Disorders Research Award to Dr. Zarate. Suported by the NIMH Intramural Research Program. FIDMAG: This work was supported by the Generalitat de Catalunya (2014 SGR 1573) and Instituto de Salud Carlos III (CPII16/00018) and (PI14/01151 and PI14/01148). PFC was supported by a Sara Borrell grant (CD19/00149, Instituto de Salud Carlos III) and a fellowship from "la Caixa" Foundation (ID 100010434, fellowship code LCF/BQ/PR22/11920017). FOR2107-Marburg: This work was funded by the German Research Foundation (DFG grant NE 2254/1-2 to Igor Nenadic, FOR2107, KI588/14-1 and FOR2107, KI588/14-2 to Tilo Kircher, Marburg, Germany). FOR2107-Muenster: This work was funded by the German Research Foundation (DFG), Udo Dannlowski (co-speaker FOR2107, DA 1151/5-1, DA 1151/5-2, grant DA1151/9-1, DA1151/10-1 and DA1151/11-1) and the Interdisciplinary Center for Clinical Research (IZKF) of the medical faculty of Münster (grant Dan3/022/22 to UD). Houston: JCS has received research grants from Compass, Alkermes, and Allergan. Melbourne: This work was supported by National Health and Medical Research Council of Australia (NHMRC) Project Grants (1064643) to BJH and to CGD (1024570). Minnesota: The study was funded by the National Institute of Mental Health (K23MH090421), the National Alliance for Research on Schizophrenia and Depression, the University of Minnesota Graduate School, the Minnesota Medical Foundation, and the Biotechnology Research Center (P41 RR008079 to the Center for Magnetic Resonance Research), University of Minnesota, and the Deborah E. Powell Center for Women's Health Seed Grant, University of Minnesota. Modal dilemma: This study was supported by the Brain and Behavior Research Foundation and by the National Health and Medical Research Council ID 1125504 to SLW. Munster: This work was funded by the German Research Foundation (SFB-TRR58, Project C09 to UD). NESDA: This study was supported by the Brain and Behavior Research Foundation and by the National Health and Medical Research Council ID 1125504 to SLW. QTIM: The QTIM dataset was supported by the Australian National Health and Medical Research Council (Project Grants No. 496682 and 1009064) and US National Institute of Child Health and Human Development (R01 HD050735). Singapore: KS was supported by National Healthcare Group, Singapore (SIG/15012) for the project. Stanford: IHG is supported in part by National Institutes of Health (R37MH101495). MDS is supported by the National Institute of Mental Health (Project Number R01MH125850), Dimension Giving Fund, Ad Astra Chandaria Foundation, Brain and Behavior Research Foundation (Grant Number 28972), BIAL Foundation (Grant Number 099/2020), Emergence Benefactors, The Ride for Mental Health, and Gatto Foundation. USFC, TIGER: UCSF Weill Institute for Neurosciences Weill Award for Investigators in the Neurosciences Impacted by COVID-19 Setbacks to TTY; this work was supported by the National Center for Complementary and Integrative Health (NCCIH) R21AT009173, R61AT009864, and 4R33AT009864-03 to TTY, by the National Center for Advancing Translational Sciences (CTSI), National Institutes of Health, through UCSF-CTSI UL1TR001872 to TTY, by the American Foundation for Suicide Prevention (AFSP) SRG-1-141-18 to TTY, by UCSF Research Evaluation and Allocation Committee (REAC) and J. Jacobson Fund to TTY, by the National Institute of Mental Health (NIMH) R01MH085734 and by the Brain and Behavior Research Foundation (formerly NARSAD) to TTY. Support for the TIGER study includes the Klingenstein Third Generation Foundation the National Institute of Mental Health K01MH117442, the Stanford Maternal Child Health Research Institute and the Stanford Center for Cognitive and Neurobiological Imaging; TCH receives partial support from the Ray and Dagmar Dolby Family Fund. TCH: This work is supported in part by the National Institute of Mental Health (K01MH117442), the Klingenstein Third Generation Foundation, Stanford Maternal Child Research Institute (Early Career Award, and K Support Award), the UCSF Research Evaluation and Allocation Committee (REAC), the Raschen-Tiedeman Fund, and the Moffitt Memorial Fund.

## References

Ayyash, S., Davis, A.D., Alders, G.L., MacQueen, G., Strother, S.C., Hassel, S., Zamyadi, M., Arnott, S.R., Harris, J.K., Lam, R.W., Milev, R., Müller, D.J., Kennedy, S.H., Rotzinger, S., Frey, B.N., Minuzzi, L., Hall, G.B., Team, C.-B.I., 2021. Exploring brain connectivity changes in major depressive disorder using functional-structural data fusion: A CAN-BIND-1 study. Hum. Brain Mapp. 42, 4940–4957. https://doi.org/10.1002/hbm.25590

Bayer, J.M.M., Thompson, P.M., Ching, C.R.K., Liu, M., Chen, A., Panzenhagen, A.C., Jahanshad, N., Marquand, A., Schmaal, L., Sämann, P.G., 2022. Site effects how-to and when: An overview of retrospective techniques to accommodate site effects in multi-site neuroimaging analyses. Front. Neurol. 13.

Belov, V., Erwin-Grabner, T., Gonul, A.S., Amod, A.R., Ojha, A., Aleman, A., Dols, A., Scharntee, A., Uyar-Demir, A., Harrison, B.J., Irungu, B.M., Besteher, B., Klimes-Dougan, B., Penninx, B.W.J.H., Mueller, B.A., Zarate, C., Davey, C.G., Ching, C.R.K., Connolly, C.G., Fu, C.H.Y., Stein, D.J., Dima, D., Linden, D.E.J., Mehler, D.M.A., Pomarol-Clotet, E., Pozzi, E., Melloni, E., Benedetti, F., MacMaster, F.P., Grabe, H.J., Völzke, H., Gotlib, I.H., Soares, J.C., Evans, J.W., Sim, K., Wittfeld, K., Cullen, K., Reneman, L., Oudega, M.L., Wright, M.J., Portella, M.J., Sacchet, M.D., Li, M., Aghajani, M., Wu, M.-J., Jaworska, N., Jahanshad, N., van der Wee, N.J.A., Groenewold, N., Hamilton, P.J., Saemann, P., Bülow, R., Poletti, S., Whittle, S., Thomopoulos, S.I., Van, S.J.A., Werff, D., Koopowitz, S.-M., Lancaster, T., Ho, T.C., Yang, T.T., Basgoze, Z., Veltman, D.J., Schmaal, L., Thompson, P.M., Goya-Maldonado, R., 2022. Multi-site benchmark classification of major depressive disorder using machine learning on cortical and subcortical measures.
Caruana, R., 1997. Multitask Learning. Mach. Learn. 28, 41–75. https://doi.org/10.1023/A:1007379606734
Coffey, C.E., Wilkinson, W.E., Weiner, R.D., Parashos, Ioanis A., Djang, W.T., Webb, M.C., Figiel, G.S., Spritzer, C.E., 1993. Quantitative Cerebral Anatomy in Depression: A Controlled Magnetic Resonance Imaging Study. Arch. Gen. Psychiatry 50, 7–16. https://doi.org/10.1001/archpsyc.1993.01820130009002
Cruz, R.S., Lebrat, L., Bourgeat, P., Fookes, C., Fripp, J., Salvado, O., 2021. DeepCSR: A 3D Deep Learning Approach for Cortical Surface Reconstruction. Presented at the Proceedings of the IEEE/CVF Winter Conference on Applications of Computer Vision, pp. 806–815.
Demirci, N., Holland, M.A., 2022. Cortical thickness systematically varies with curvature and depth in healthy human brains. Hum. Brain Mapp. 43, 2064–2084. https://doi.org/10.1002/hbm.25776
Deng, J., Dong, W., Socher, R., Li, L.-J., Li, K., Fei-Fei, L., 2009. ImageNet: A large-scale hierarchical image database, in: 2009 IEEE Conference on Computer Vision and Pattern Recognition. Presented at the 2009 IEEE Conference on Computer Vision and Pattern Recognition, pp. 248–255. https://doi.org/10.1109/CVPR.2009.5206848
Depping, M.S., Thomann, P.A., Wolf, N.D., Vasic, N., Sosic-Vasic, Z., Schmitgen, M.M., Sambataro, F., Wolf, R.C., 2018. Common and distinct patterns of abnormal cortical gyrification in major depression and borderline personality disorder. Eur. Neuropsychopharmacol. 28, 1115–1125. https://doi.org/10.1016/j.euroneuro.2018.07.100
Dinga, R., Penninx, B., Veltman, D., Schmaal, L., Marquand, A., 2019. Beyond accuracy: Measures for assessing machine learning models, pitfalls and guidelines. https://doi.org/10.1101/743138
Dinsdale, N., Jenkinson, M., Namburete, A., 2020. Deep Learning-Based Unlearning of Dataset Bias for MRI Harmonisation and Confound Removal. NeuroImage 228, 117689. https://doi.org/10.1016/j.neuroimage.2020.117689
Flint, C., Cearns, M., Opel, N., Redlich, R., Mehler, D.M.A., Emden, D., Winter, N.R., Leenings, R., Eickhoff, S.B., Kircher, T., Krug, A., Nenadic, I., Arolt, V., Clark, S., Baune, B.T., Jiang, X., Dannlowski, U., Hahn, T., 2021. Systematic misestimation of machine learning performance in neuroimaging studies of depression. Neuropsychopharmacology 1–8. https://doi.org/10.1038/s41386-021-01020-7
Fornito, A., Zalesky, A., Breakspear, M., 2015. The connectomics of brain disorders. Nat. Rev. Neurosci. 16, 159–172. https://doi.org/10.1038/nrn3901
Friedrich, M.J., 2017. Depression Is the Leading Cause of Disability Around the World. JAMA 317, 1517. https://doi.org/10.1001/jama.2017.3826
Gao, K., Fan, Z., Su, J., Zeng, L.-L., Shen, H., Zhu, J., Hu, D., 2021. Deep Transfer Learning for Cerebral Cortex Using Area-Preserving Geometry Mapping. Cereb. Cortex. https://doi.org/10.1093/cercor/bhab394
Gao, S., Calhoun, V.D., Sui, J., 2018. Machine learning in major depression: From classification to treatment outcome prediction. CNS Neurosci. Ther. 24, 1037–1052. https://doi.org/10.1111/cns.13048
Henschel, L., Conjeti, S., Estrada, S., Diers, K., Fischl, B., Reuter, M., 2020. FastSurfer - A fast and accurate deep learning based neuroimaging pipeline. NeuroImage 219, 117012. https://doi.org/10.1016/j.neuroimage.2020.117012
Hettwer, M.D., Larivière, S., Park, B.Y., van den Heuvel, O.A., Schmaal, L., Andreassen, O.A., Ching, C.R.K., Hoogman, M., Buitelaar, J., van Rooij, D., Veltman, D.J., Stein, D.J., Franke, B., van Erp, T.G.M., Jahanshad, N., Thompson, P.M., Thomopoulos, S.I., Bethlehem, R. a. I., Bernhardt, B.C., Eickhoff, S.B., Valk, S.L., 2022. Coordinated cortical thickness alterations across six neurodevelopmental and psychiatric disorders. Nat. Commun. 13, 6851. https://doi.org/10.1038/s41467-022-34367-6

Ho, T.C., Gutman, B., Pozzi, E., Grabe, H.J., Hosten, N., Wittfeld, K., Völzke, H., Baune, B., Dannlowski, U., Förster, K., Grotegerd, D., Redlich, R., Jansen, A., Kircher, T., Krug, A., Meinert, S., Nenadic, I., Opel, N., Dinga, R., Veltman, D.J., Schnell, K., Veer, I., Walter, H., Gotlib, I.H., Sacchet, M.D., Aleman, A., Groenewold, N.A., Stein, D.J., Li, M., Walter, M., Ching, C.R.K., Jahanshad, N., Ragothaman, A., Isaev, D., Zavaliangos-Petropulu, A., Thompson, P.M., Sämann, P.G., Schmaal, L., 2022. Subcortical shape alterations in major depressive disorder: Findings from the ENIGMA major depressive disorder working group. Hum. Brain Mapp. 43, 341–351. https://doi.org/10.1002/hbm.24988

Ho, T.C., Sisk, L.M., Kulla, A., Teresi, G.I., Hansen, M.M., Wu, H., Gotlib, I.H., 2021. Sex differences in myelin content of white matter tracts in adolescents with depression. Neuropsychopharmacology 46, 2295–2303. https://doi.org/10.1038/s41386-021-01078-3

Hsu, C., Chang, C., Lin, C.-J., 2003. A Practical Guide to Support Vector Classification Chih-Wei Hsu, Chih-Chung Chang, and Chih-Jen Lin.

Huang, G., Liu, Z., Van Der Maaten, L., Weinberger, K.Q., 2017. Densely Connected Convolutional Networks, in: 2017 IEEE Conference on Computer Vision and Pattern Recognition (CVPR). Presented at the 2017 IEEE Conference on Computer Vision and Pattern Recognition (CVPR), pp. 2261–2269. https://doi.org/10.1109/CVPR.2017.243

Kambeitz, J., Cabral, C., Sacchet, M.D., Gotlib, I.H., Zahn, R., Serpa, M.H., Walter, M., Falkai, P., Koutsouleris, N., 2017. Detecting Neuroimaging Biomarkers for Depression: A Meta-analysis of Multivariate Pattern Recognition Studies. Biol. Psychiatry 82, 330–338. https://doi.org/10.1016/j.biopsych.2016.10.028

Kohoutová, L., Heo, J., Cha, S., Lee, S., Moon, T., Wager, T.D., Woo, C.-W., 2020. Toward a unified framework for interpreting machine-learning models in neuroimaging. Nat. Protoc. 15, 1399–1435. https://doi.org/10.1038/s41596-019-0289-5

Kremen, W.S., Prom-Wormley, E., Panizzon, M.S., Eyler, L.T., Fischl, B., Neale, M.C., Franz, C.E., Lyons, M.J., Pacheco, J., Perry, M.E., Stevens, A., Schmitt, J.E., Grant, M.D., Seidman, L.J., Thermenos, H.W., Tsuang, M.T., Eisen, S.A., Dale, A.M., Fennema-Notestine, C., 2010. Genetic and environmental influences on the size of specific brain regions in midlife: the VETSA MRI study. NeuroImage 49, 1213–1223. https://doi.org/10.1016/j.neuroimage.2009.09.043

Lecun, Y., Bottou, L., Bengio, Y., Haffner, P., 1998. Gradient-based learning applied to document recognition. Proc. IEEE 86, 2278–2324. https://doi.org/10.1109/5.726791

Li, J., Seidlitz, J., Suckling, J., Fan, F., Ji, G.-J., Meng, Y., Yang, S., Wang, K., Qiu, J., Chen, H., Liao, W., 2021. Cortical structural differences in major depressive disorder correlate with cell type-specific transcriptional signatures. Nat. Commun. 12, 1647. https://doi.org/10.1038/s41467-021-21943-5

Lu, X., Yang, Y., Wu, F., Gao, M., Xu, Y., Zhang, Y., Yao, Y., Du, X., Li, C., Wu, L., Zhong, X., Zhou, Y., Fan, N., Zheng, Y., Xiong, D., Peng, H., Escudero, J., Huang, B., Li, X., Ning, Y., Wu, K., 2016. Discriminative analysis of schizophrenia using support vector machine and recursive feature elimination on structural MRI images. Medicine (Baltimore) 95, e3973. https://doi.org/10.1097/MD.0000000000003973

Ma, Q., Zhang, T., Zanetti, M.V., Shen, H., Satterthwaite, T.D., Wolf, D.H., Gur, R.E., Fan, Y., Hu, D., Busatto, G.F., Davatzikos, C., 2018. Classification of multi-site MR images in the presence of heterogeneity using multi-task learning. NeuroImage Clin. 19, 476–486. https://doi.org/10.1016/j.nicl.2018.04.037

Machado, M., Iskedjian, M., Ruiz, I., Einarson, T.R., 2006. Remission, dropouts, and adverse drug reaction rates in major depressive disorder: a meta-analysis of head-to-head trials. Curr. Med. Res. Opin. 22, 1825–1837. https://doi.org/10.1185/030079906X132415

Mårtensson, G., Ferreira, D., Granberg, T., Cavallin, L., Oppedal, K., Padovani, A., Rektorova, I., Bonanni, L., Pardini, M., Kramberger, M.G., Taylor, J.-P., Hort, J., Snædal, J., Kulisevsky, J., Blanc, F., Antonini, A., Mecocci, P., Vellas, B., Tsolaki, M., Kłoszewska, I., Soininen, H., Lovestone, S., Simmons, A., Aarsland, D., Westman, E., 2020. The reliability of a deep learning model in clinical out-of-distribution MRI data: A multicohort study. Med. Image Anal. 66, 101714. https://doi.org/10.1016/j.media.2020.101714
Mendlewicz, J., 2008. Towards achieving remission in the treatment of depression. Dialogues Clin. Neurosci. 10, 371–375.
Nakano, T., Takamura, M., Ichikawa, N., Okada, G., Okamoto, Y., Yamada, M., Suhara, T., Yamawaki, S., Yoshimoto, J., 2020. Enhancing Multi-Center Generalization of Machine Learning-Based Depression Diagnosis From Resting-State fMRI. Front. Psychiatry 11.
Pinaya, W.H.L., Mechelli, A., Sato, J.R., 2018. Using deep autoencoders to identify abnormal brain structural patterns in neuropsychiatric disorders: A large-scale multi-sample study. Hum. Brain Mapp. 40, 944–954. https://doi.org/10.1002/hbm.24423
Pomponio, R., Erus, G., Habes, M., Doshi, J., Srinivasan, D., Mamourian, E., Bashyam, V., Nasrallah, I.M., Satterthwaite, T.D., Fan, Y., Launer, L.J., Masters, C.L., Maruff, P., Zhuo, C., Völzke, H., Johnson, S.C., Fripp, J., Koutsouleris, N., Wolf, D.H., Gur, Raquel, Gur, Ruben, Morris, J., Albert, M.S., Grabe, H.J., Resnick, S.M., Bryan, R.N., Wolk, D.A., Shinohara, R.T., Shou, H., Davatzikos, C., 2020. Harmonization of large MRI datasets for the analysis of brain imaging patterns throughout the lifespan. NeuroImage 208, 116450. https://doi.org/10.1016/j.neuroimage.2019.116450
Qin, K., Lei, D., Pinaya, W.H.L., Pan, N., Li, W., Zhu, Z., Sweeney, J.A., Mechelli, A., Gong, Q., 2022. Using graph convolutional network to characterize individuals with major depressive disorder across multiple imaging sites. eBioMedicine 78, 103977. https://doi.org/10.1016/j.ebiom.2022.103977
Radua, J., Vieta, E., Shinohara, R., Kochunov, P., Quidé, Y., Green, M.J., Weickert, C.S., Weickert, T., Bruggemann, J., Kircher, T., Nenadić, I., Cairns, M.J., Seal, M., Schall, U., Henskens, F., Fullerton, J.M., Mowry, B., Pantelis, C., Lenroot, R., Cropley, V., Loughland, C., Scott, R., Wolf, D., Satterthwaite, T.D., Tan, Y., Sim, K., Piras, Fabrizio, Spalletta, G., Banaj, N., Pomarol-Clotet, E., Solanes, A., Albajes-Eizagirre, A., Canales-Rodríguez, E.J., Sarro, S., Di Giorgio, A., Bertolino, A., Stäblein, M., Oertel, V., Knöchel, C., Borgwardt, S., du Plessis, S., Yun, J.Y., Kwon, J.S., Dannlowski, U., Hahn, T., Grotegerd, D., Alloza, C., Arango, C., Janssen, J., Díaz-Caneja, C., Jiang, W., Calhoun, V., Ehrlich, S., Yang, K., Cascella, N.G., Takayanagi, Y., Sawa, A., Tomyshev, A., Lebedeva, I., Kaleda, V., Kirschner, M., Hoschl, C., Tomecek, D., Skoch, A., van Amelsvoort, T., Bakker, G., James, A., Preda, A., Weideman, A., Stein, D.J., Howells, F., Uhlmann, A., Temmingh, H., López-Jaramillo, C., Díaz-Zuluaga, A., Fortea, L., Martinez-Heras, E., Solana, E., Llufriu, S., Jahanshad, N., Thompson, P., Turner, J., van Erp, T., Glahn, D., Pearlson, G., Hong, E., Krug, A., Carr, V., Tooney, P., Cooper, G., Rasser, P., Michie, P., Catts, S., Gur, Raquel, Gur, Ruben, Yang, F., Fan, F., Chen, J., Guo, H., Tan, S., Wang, Z., Xiang, H., Piras, Federica, Assogna, F., Salvador, R., McKenna, P., Bonvino, A., King, M., Kaiser, S., Nguyen, D., Pineda-Zapata, J., 2020. Increased power by harmonizing structural MRI site differences with the ComBat batch adjustment method in ENIGMA. NeuroImage 218. https://doi.org/10.1016/j.neuroimage.2020.116956
Rush, A.J., Trivedi, M.H., Wisniewski, S.R., Nierenberg, A.A., Stewart, J.W., Warden, D., Niederehe, G., Thase, M.E., Lavori, P.W., Lebowitz, B.D., McGrath, P.J., Rosenbaum, J.F., Sackeim, H.A., Kupfer, D.J., Luther, J., Fava, M., 2006. Acute and longer-term outcomes in depressed outpatients requiring one or several treatment steps: a STAR*D report. Am. J. Psychiatry 163, 1905–1917. https://doi.org/10.1176/ajp.2006.163.11.1905
Sacchet, M.D., Gotlib, I.H., 2017. Myelination of the brain in Major Depressive Disorder: An in vivo quantitative magnetic resonance imaging study. Sci. Rep. 7, 2200. https://doi.org/10.1038/s41598-017-02062-y
Sacchet, M.D., Prasad, G., Foland-Ross, L.C., Thompson, P.M., Gotlib, I.H., 2015. Support vector machine classification of major depressive disorder using diffusion-weighted neuroimaging and graph theory. Front. Psychiatry 6, 1–10. https://doi.org/10.3389/fpsyt.2015.00021
Schmaal, L., Hibar, D.P., Sämann, P.G., Hall, G.B., Baune, B.T., Jahanshad, N., Cheung, J.W., Van Erp, T.G.M., Bos, D., Ikram, M.A., Vernooij, M.W., Niessen, W.J., Tiemeier, H., Hofman, A.,

Wittfeld, K., Grabe, H.J., Janowitz, D., Bülow, R., Selonke, M., Völzke, H., Grotegerd, D., Dannlowski, U., Arolt, V., Opel, N., Heindel, W., Kugel, H., Hoehn, D., Czisch, M., Couvy-Duchesne, B., Rentería, M.E., Strike, L.T., Wright, M.J., Mills, N.T., De Zubicaray, G.I., McMahon, K.L., Medland, S.E., Martin, N.G., Gillespie, N.A., Goya-Maldonado, R., Gruber, O., Krämer, B., Hatton, S.N., Lagopoulos, J., Hickie, I.B., Frodl, T., Carballedo, A., Frey, E.M., Van Velzen, L.S., Penninx, B.W.J.H., Van Tol, M.J., Van der Wee, N.J., Davey, C.G., Harrison, B.J., Mwangi, B., Cao, B., Soares, J.C., Veer, I.M., Walter, H., Schoepf, D., Zurowski, B., Konrad, C., Schramm, E., Normann, C., Schnell, K., Sacchet, M.D., Gotlib, I.H., MacQueen, G.M., Godlewska, B.R., Nickson, T., McIntosh, A.M., Papmeyer, M., Whalley, H.C., Hall, J., Sussmann, J.E., Li, M., Walter, M., Aftanas, L., Brack, I., Bokhan, N.A., Thompson, P.M., Veltman, D.J., 2017. Cortical abnormalities in adults and adolescents with major depression based on brain scans from 20 cohorts worldwide in the ENIGMA Major Depressive Disorder Working Group. Mol. Psychiatry 22, 900–909. https://doi.org/10.1038/mp.2016.60

Schmaal, L., Veltman, D.J., van Erp, T.G.M., Sämann, P.G., Frodl, T., Jahanshad, N., Loehrer, E., Tiemeier, H., Hofman, A., Niessen, W.J., Vernooij, M.W., Ikram, M.A., Wittfeld, K., Grabe, H.J., Block, A., Hegenscheid, K., Völzke, H., Hoehn, D., Czisch, M., Lagopoulos, J., Hatton, S.N., Hickie, I.B., Goya-Maldonado, R., Krämer, B., Gruber, O., Couvy-Duchesne, B., Rentería, M.E., Strike, L.T., Mills, N.T., de Zubicaray, G.I., McMahon, K.L., Medland, S.E., Martin, N.G., Gillespie, N.A., Wright, M.J., Hall, G.B., MacQueen, G.M., Frey, E.M., Carballedo, A., van Velzen, L.S., van Tol, M.J., van der Wee, N.J., Veer, I.M., Walter, H., Schnell, K., Schramm, E., Normann, C., Schoepf, D., Konrad, C., Zurowski, B., Nickson, T., McIntosh, A.M., Papmeyer, M., Whalley, H.C., Sussmann, J.E., Godlewska, B.R., Cowen, P.J., Fischer, F.H., Rose, M., Penninx, B.W.J.H., Thompson, P.M., Hibar, D.P., 2016. Subcortical brain alterations in major depressive disorder: findings from the ENIGMA Major Depressive Disorder working group. Mol. Psychiatry 21, 806–812. https://doi.org/10.1038/mp.2015.69

Schulz, M.-A., Yeo, B.T.T., Vogelstein, J.T., Mourao-Miranada, J., Kather, J.N., Kording, K., Richards, B., Bzdok, D., 2020. Different scaling of linear models and deep learning in UKBiobank brain images versus machine-learning datasets. Nat. Commun. 11, 1–15. https://doi.org/10.1038/s41467-020-18037-z

Seong, S.-B., Pae, C., Park, H.-J., 2018. Geometric Convolutional Neural Network for Analyzing Surface-Based Neuroimaging Data. Front. Neuroinformatics 12. https://doi.org/10.3389/fninf.2018.00042

Sheline, Y.I., Gado, M.H., Kraemer, H.C., 2003. Untreated depression and hippocampal volume loss. Am. J. Psychiatry 160, 1516–1518. https://doi.org/10.1176/appi.ajp.160.8.1516

Solanes, A., Gosling, C.J., Fortea, L., Ortuño, M., Lopez-Soley, E., Llufriu, S., Madero, S., Martinez-Heras, E., Pomarol-Clotet, E., Solana, E., Solé, E., Vieta, E., Radua, J., 2022. Removing the effects of the site in brain imaging machine-learning – Measurement and extendable benchmark. NeuroImage 119800. https://doi.org/10.1016/j.neuroimage.2022.119800

Solanes, A., Palau, P., Fortea, L., Salvador, R., González-Navarro, L., Llach, C.D., Valentí, M., Vieta, E., Radua, J., 2021. Biased accuracy in multisite machine-learning studies due to incomplete removal of the effects of the site. Psychiatry Res. Neuroimaging 314, 111313. https://doi.org/10.1016/j.pscychresns.2021.111313

Stolicyn, A., Harris, M.A., Shen, X., Barbu, M.C., Adams, M.J., Hawkins, E.L., de Nooij, L., Yeung, H.W., Murray, A.D., Lawrie, S.M., Steele, J.D., McIntosh, A.M., Whalley, H.C., 2020. Automated classification of depression from structural brain measures across two independent community-based cohorts. Hum. Brain Mapp. 41, 3922–3937. https://doi.org/10.1002/hbm.25095

Su, Z., Wang, Y., Shi, R., Zeng, W., Sun, J., Luo, F., Gu, X., 2015. Optimal Mass Transport for Shape Matching and Comparison. IEEE Trans. Pattern Anal. Mach. Intell. 37, 2246–2259. https://doi.org/10.1109/TPAMI.2015.2408346

van Velzen, L., Kelly, S., Isaev, D., Aleman, A., Aftanas, L., Bauer, J., Baune, B., Brak, I., Carballedo, A., Connolly, C., Couvy-Duchesne, B., Cullen, K., Danilenko, K., Dannlowski, U., Enneking, V.,

Filimonova, E., Förster, K., Frodl, T., Gotlib, I., Schmaal, L., 2020. White matter disturbances in major depressive disorder: a coordinated analysis across 20 international cohorts in the ENIGMA MDD working group. Mol. Psychiatry 25. https://doi.org/10.1038/s41380-019-0477-2

Wen, J., Thibeau-Sutre, E., Diaz-Melo, M., Samper-González, J., Routier, A., Bottani, S., Dormont, D., Durrleman, S., Burgos, N., Colliot, O., 2020. Convolutional neural networks for classification of Alzheimer's disease: Overview and reproducible evaluation. Med. Image Anal. 63, 101694. https://doi.org/10.1016/j.media.2020.101694

White, T., Andreasen, N.C., Nopoulos, P., 2002. Brain volumes and surface morphology in monozygotic twins. Cereb. Cortex N. Y. N 1991 12, 486–493. https://doi.org/10.1093/cercor/12.5.486

Wig, G., Laumann, T., Cohen, A., Power, J., Nelson, S., Glasser, M., Miezin, F., Snyder, A., Schlaggar, B., Petersen, S., 2013. Parcellating an Individual Subject's Cortical and Subcortical Brain Structures Using Snowball Sampling of Resting-State Correlations. Cereb. Cortex N. Y. N 1991 24. https://doi.org/10.1093/cercor/bht056

Williams, L.Z.J., Fawaz, A., Glasser, M.F., Edwards, A.D., Robinson, E.C., 2021. Geometric Deep Learning of the Human Connectome Project Multimodal Cortical Parcellation, in: Abdulkadir, A., Kia, S.M., Habes, M., Kumar, V., Rondina, J.M., Tax, C., Wolfers, T. (Eds.), Machine Learning in Clinical Neuroimaging, Lecture Notes in Computer Science. Springer International Publishing, Cham, pp. 103–112. https://doi.org/10.1007/978-3-030-87586-2_11

Winter, N.R., Leenings, R., Ernsting, J., Sarink, K., Fisch, L., Emden, D., Blanke, J., Goltermann, J., Opel, N., Barkhau, C., Meinert, S., Dohm, K., Repple, J., Mauritz, M., Gruber, M., Leehr, E.J., Grotegerd, D., Redlich, R., Jansen, A., Nenadic, I., Nöthen, M., Forstner, A., Rietschel, M., Groß, J., Bauer, J., Heindel, W., Andlauer, T., Eickhoff, S., Kircher, T., Dannlowski, U., Hahn, T., 2021. More Alike than Different: Quantifying Deviations of Brain Structure and Function in Major Depressive Disorder across Neuroimaging Modalities. ArXiv211210730 Q-Bio.

Wottschel, V., Chard, D.T., Enzinger, C., Filippi, M., Frederiksen, J.L., Gasperini, C., Giorgio, A., Rocca, M.A., Rovira, A., De Stefano, N., Tintoré, M., Alexander, D.C., Barkhof, F., Ciccarelli, O., MAGNIMS study group and the EuroPOND consortium, 2019. SVM recursive feature elimination analyses of structural brain MRI predicts near-term relapses in patients with clinically isolated syndromes suggestive of multiple sclerosis. NeuroImage Clin. 24, 102011. https://doi.org/10.1016/j.nicl.2019.102011

Yan, W., Calhoun, V., Song, M., Cui, Y., Yan, H., Liu, S., Fan, L., Zuo, N., Yang, Z., Xu, K., Yan, J., Lv, L., Chen, J., Chen, Y., Guo, H., Li, P., Lu, L., Wan, P., Wang, Huaning, Wang, Huiling, Yang, Y., Zhang, H., Zhang, D., Jiang, T., Sui, J., 2019. Discriminating schizophrenia using recurrent neural network applied on time courses of multi-site FMRI data. EBioMedicine 47, 543–552. https://doi.org/10.1016/j.ebiom.2019.08.023

Zhang, X., Yao, S., Zhu, Xiongzhao, Wang, X., Zhu, Xueling, Zhong, M., 2012. Gray matter volume abnormalities in individuals with cognitive vulnerability to depression: A voxel-based morphometry study. J. Affect. Disord. 136, 443–452. https://doi.org/10.1016/j.jad.2011.11.005

Zhang, Y., Jenkins, D.F., Manimaran, S., Johnson, W.E., 2018. Alternative empirical Bayes models for adjusting for batch effects in genomic studies. BMC Bioinformatics 19, 262. https://doi.org/10.1186/s12859-018-2263-6

Zhang, Y., Yu, C., Zhou, Y., Li, K., Li, C., Jiang, T., 2009. Decreased gyrification in major depressive disorder. Neuroreport 20, 378–380. https://doi.org/10.1097/WNR.0b013e3283249b34

Zhang, Y.-D., Dong, Z., Wang, S.-H., Yu, X., Yao, X., Zhou, Q., Hu, H., Li, M., Jiménez-Mesa, C., Ramirez, J., Martinez, F.J., Gorriz, J.M., 2020. Advances in multimodal data fusion in neuroimaging: Overview, challenges, and novel orientation. Inf. Fusion 64, 149–187. https://doi.org/10.1016/j.inffus.2020.07.006

Zhou, J., Cui, G., Zhang, Z., Yang, C., Liu, Z., Wang, L., Li, C., Sun, M., 2019. Graph Neural Networks: A Review of Methods and Applications. ArXiv181208434 Cs Stat.

## Supplementary Materials for

# Classification of Major Depressive Disorder Using Vertex-Wise Brain Sulcal Depth, Curvature, and Thickness with a Deep and a Shallow Learning Model

**Supplementary Table 1:** ENIGMA-MDD Instrument for diagnosing major depressive disorder (MDD) and exclusion criteria by site

| Cohort | Diagnosis measurment | Sample characteristics/Inclusion criteria | Exclusion criteria |
|---|---|---|---|
| **AFFDIS** | ICD-10/DSM-IV criteria | MDD subjects currently depressed and in day program or inpatient | All subjects exclusion criteria: current or history of neurological disorder or brain injury, current substance abuse or dependence (not including nicotine), pregnancy, MRI contraindications, inability to give consent. MDD specific: comorbid psychiatric diagnosis. Healthy control specific: current or history of psychiatric diagnosis. |
| **Pharmo (AMC)** | MINI Plus | 48 subjects with lifetime diagnosis of either MDD and/or AD and 14 healthy controls. Patients were startified depending on exposure to SSRIs: early (before age 23) or late (after age 23) exposure to SSRI's, or no exposure at all (UN). 15 subjects were diagnosed with only MDD, 3 with only AD and 22 with both MDD and AD (8 subjects did not receive a diagnosis due to incomplete M.I.N.I. Plus assessment). According to the M.I.N.I. Plus, none of the HC subjects were ever diagnosed with MDD or AD | Less than three week medication-free interval before scanning, current psychotropic medication use, a history of chronic or neurological disorder, family history of sudden heart failure or epileptic attacks, pregnancy (tested via urine sampling prior to the assessment), breast feeding, alcohol dependence and contra-indications for an MRI scan (e.g., ferromagnetic fragments). Participants agreed to abstain from smoking, caffeine and alcohol use for 24 hours prior to the assessments. |
| **Barcelona** | DSM-IV-TR acc. to CIDI-interview and HAMD | Outpatients with MDD diagnosis (DSM-IV-TR), with a first episode, recurrent MDD or chronic MDD (TRD) age 18-65 | The exclusion criteria for healthy participants were: lifetime psychiatric diagnoses, first-degree relatives with psychiatric diagnoses and clinically significant physical or neurological illnesses. Axis I comorbidity according to DSM-IV-TR criteria was an exclusion criteria for all participants. |
| **Cardiff** | Hamilton Depression Rating Scale (HDRS-17) | N= 40, MDD patients with a current moderate to severe depressive episode despite minimum three months of stable antidepressant treatment | Psychotic symptoms, current substance dependence, eating disorders, claustrophobia and other MRI contraindications, and ongoing non-pharmacological treatment. |
| **CSAN (Adf)** | MINI | Current MDD: Meets MINI criteria for depression; comorbid anxiety disorders are allowed; mood-congruent psychotic symptoms allowed. | Current MDD: a current DSM-5 diagnosis of substance use disorder, except nicotine; a psychotic disorder, except depression with mood-congruent psychotic features; new antidepressant medication during the month before study participation (two months for fluoxetine); change of the dose of psychotropic medications over the last month (antidepressant and antipsychotic medication) |
| | | | or the last two months (mood stabilizers and anticonvulsants). |
| **Calgary** | KSADS | First episode MDD and healthy controls (Dalhousie sample). Recurrent MDD and healthy controls, recruited via referral from clinicians in Calgary, Alberta and through advertisements in local clinics and at the University of Calgary (Calgary sample). | Dalhousie Sample: A history of neurological illness, medical illness, claustrophobia, >21 year of age, or the presence of a ferrous implant or pacemaker. University of Calgary: Left handed; history of seizures, epilepsy or other neurological or psychiatric diagnoses (specifically bipolar disorder, psychosis, pervasive developmental disorder, eating disorders, PTSD); pregnancy |
| **DCHS** | MINI | Women over the age of 18 years, who were between 20 and 28 weeks pregnant, who presented at either of the two recruitment clinics, and who had no intention of moving out of the area within the following year, and were able to give written consent | 1) loss of consciousness longer than 30 minutes, 2) inability to speak English, 3) current/lifetime alcohol and/or substance dependence or abuse, 4) psychopathology other than PTSD and/or MDD, 5) traumatic brain injury, 6) standard MRI exclusion criteria |
| **FIDMAG** | DSM-IV-TR criteria | MDD patients within a current depressive episode (HDRS >= 17, only 1 patient was in remission), right-handed, age 18-65 | Patients were excluded (i) if they were left-handed; (ii) if they were younger than 18 or older than 65 years; (iii) if they had a history of brain trauma or neurological disease; (iv) if they had shown alcohol/ substance abuse within 12 months prior to participation; and (v) if they had undergone electroconvulsive therapy in the previous 12 months. |
| **FOR2107Marburg** | SCID-1 | Participants recruited by means of public advertisement and from the inpatient services. Inclusion criteria: age 18-65 years; patients were diagnosed with major depressive disorder by SCID-Interview, currently depressed or remitted. | Exclusion criteria all: any MRI contraindications; any neurological abnormalities. Exclusion criteria controls: any current or former psychiatric disorder; Exclusion criteria patients: substance dependence or current benzodiazepine treatment (wash out of at least three half-lives before study participation)" |
| **FOR2107Munster** | SCID-1 | Participants recruited by means of public advertisement and from the inpatient services. Inclusion criteria: age 18-65 years; patients were diagnosed with major depressive disorder by SCID-Interview, currently depressed or remitted. | Exclusion criteria all: any MRI contraindications; any neurological abnormalities. Exclusion criteria controls: any current or former psychiatric disorder; Exclusion criteria patients: substance dependence or current benzodiazepine treatment (wash out of at least three half-lives before study participation)" |
| **Houston** | SCID interview | Outpatients | MDD subjects: age below 18; lifetime or current diagnosis of psychotic disorder, or bipolar I or II disorder; substance abuse/dependence in 6 months prior to study inclusion; current major medical problems. Control subjects: age below 18; current major medical problems; current psychiatric or neurologic disorder; history of psychiatric disorders in a first-degree relative; current major medical problems. Both groups: MRI contra-indications |
| **Hiroshima** | MINI | MDD Patients were recruited from local clinics, 20-80 years. Controls were recruited | MDD patients: comorbid psychiatric disorders other than MDD, Control subjects: any history of psychiatric disorder |
| | | from local community by advertising in local papers. | |
| **TiPs (Jena, Germany)** | SCID interview | Psychiatric inpatients and tinnitus patients with MDD or a disorder of the depressive spectrum (also adjustment disorders as pointed out in the data table); psychiatrically healthy controls were derived from community and tinnitus patients | MDD subjects: presence of axis-I disorders other than MDD or adjustment disorders. Control subjects: no Axis-I diagnosis, no medication use. Exclusion criteria for all subjects included history of neurological disease (e.g. tumour, head trauma, epilepsy) or untreated internal medical condtitions, intellectual and/or developmental disability. Only German native speakers were allowed to participate. |
| **MODECT** | MINI | Older adults, aged above 55, with severe depression admitted to be treated with ECT | Exclusion criteria were another major DSM-IV-TR diagnosis, such as schizophrenia, bipolar or schizoaffective disorder and a history of major neurological illness (including Parkinson's disease, stroke and dementia). |
| **Melbourne** | SCID interview | Youth depression sample: 15-25 years of age. Recruited as part of 2 large RCTs (incl. YoDA-C - Davey et al., 2014; Trials) and scanned prior to treatment randomisation. 60 patients unmedicated (YoDA-C). | MDD subjects: lifetime or current SCID-I diagnosis of psychotic disorder, or bipolar I or II disorder. Control subjects: any SCID-I diagnosis or medication use. Both groups: Acute or unstable medical disorder; general MRI contraindications |
| **Minnesota** | Schedule for Affective Disorders and Schizophrenia for School-Age Children–Present and Lifetime Version and the Children's Depression Rating Scale–Revised (CDRS-R). | Adolescents with MDD and HCs aged 12 to 19 years were recruited to participate through community postings and referrals from local mental health services. Adolescents with MDD were eligible if they had a primary diagnosis of MDD and had not received any psychotropic medication treatment for the past 2 months. Healthy adolescents were eligible if they had no current or past psychiatric diagnoses and were frequency matched to the MDD group on age and sex | Exclusion criteria for both groups included the presence of a neurologic or other chronic medical condition, mental retardation, pervasive developmental disorder, substance use disorder, bipolar disorder, or schizophrenia |
| **MOODS / DEP-ARREST CLIN** | MINI, DSM5 | Patients aged 18-65 years with a current MDE diagnosis (MINI interview(Sheehan et al., 1998) and a minimum depression score of 18 on the Hamilton Depression Rating Scale-17 items (HDRS) in the context of MDD, as well as free of antidepressant drug use at least one month before the study beginning, were included. HCs were included based on the absence of current or past mental disorders or somatic conditions, particularly nasal polyposis and chronic or acute sinusitis or rhinitis | Patients suffering from bipolar disorder, psychotic disorder, eating disorder, and addictions, according to the DSM-5 criteria, or from nasal polyposis, chronic or acute sinusitis, chronic or acute rhinitis or pregnancy or breastfeeding, were not included. HCs were included based on the absence of current or past mental disorders or somatic conditions, particularly nasal polyposis and chronic or acute sinusitis or rhinitis |
| **Moral Dilemma** | SCID interview | Youth depression sample: 15-25 years of age; recruited from outpatient service. Controls recruited from general community. | MDD subjects: lifetime or current SCID-I diagnosis of psychotic disorder, or bipolar I or II disorder; current antidepressant medication use. Control subjects: any SCID-I diagnosis or medication use. Both groups: Acute or unstable medical disorder; general MRI contraindications |
| **Munster** | SCID interview | Participants recruited by means of public advertisement and from the inpatient services. Inclusion criteria: age 16-65 years; | MDD subjects: presence of bipolar disorder, schizoaffective disorders and schizophrenia; substancerelated disorders or current |
| | | patients were diagnosed with major depressive disorder by SCID-Interview | benzodiazepine treatment (wash out of at least three half-lives before study participation), and former electroconvulsive therapy. Control subjects: any current or former psychiatric disorder. Both groups: any neurological abnormalities, MRI contra-indications |
| **NESDA** | CIDI interview | DSM-4 based diagnosis of MDD (6 month recency), using CIDI interview. 93 (60%) MDD patients have a comorbid ANX diagnosis. Age range 18-65 | N/A |
| **QTIM** | CIDI interview | Retrospective questionnaire about depression episodes combined with an MRI study. The best described MDD episode is defined as the worst one (according to individuals). We have up to 5 supplementary episodes (briefly) described. Sample composed of twins and relatives. Population-based sample | MDD subjects: presence of axis-I disorders other than MDD and anxiety disorders Control subjects: antidepressant use, psychiatric disorders All subjects: relatedness between subjects, left handedness, history of neurological or other severe medical illness, head injury or current or past diagnosis of substance abuse, use of cognition affecting medication and general MRI contraindications |
| **San Francisco UCSF** | KSADS (semi-structured interview based on DSM) for MDD, DISC/DPS for HCL | Outpatient/community-based sample with DSM diagnosis, mostly antidepressant-naive and approximately 60% of MDD have comorbid anxiety disorders | Exclusion criteria for all participants included: 1) use of pharmacotherapeutics for treating psychiatric conditions within the past 6 months, 2) misuse of drugs within two months prior to MRI scanning; 3) two or more alcoholic drinks per week within the previous month (as assessed by the Customary Drinking and Drug Use Record; CDDR) (Brown et al, 1998); 4) a full scale IQ score of less than 75 (as assessed by the Wechsler Abbreviated Scale of Intelligence; WASI) (Wechsler, 1999); 5) contraindications for MRI including ferromagnetic implants and claustrophobia; 6) pregnancy or the possibility of pregnancy; 7) left-handedness; 8) prepubertal status (as assessed as Tanner stages of 1 or 2) (Tanner, 1962); 9) inability to understand and comply with procedures; 10) neurological disorder (including meningitis, migraine, or HIV); 11) head trauma; 12) learning disability; 13) serious health problems; and 14) complicated or premature birth (i.e., birth before 33 weeks of gestation). The MDD group was subject to the additional exclusion criterion of a primary psychiatric diagnosis other than MDD. The HCL group was subject to the additional exclusion criteria of: 1) history of mood or psychotic disorders in a first- or second-degree relative (as assessed by the Family Interview for Genetics; FIGS) (Maxwell, 1992); and 2) current or lifetime DSM-IV-TR Axis I psychiatric disorder. |
| **SHIP_START-2** | M-CIDI interview | Population based longitudinal cohort study | MDD subjects: presence of axis-I disorders other than MDD, anxiety disorders, conversion, somatization and eating disorder. Control subjects: no lifetime diagnosis of depression, no antidepressiva, and severity index=0 All subjects: We removed subjects with medical conditions (e.g. a history of cerebral tumor, stroke, Parkinson's diseases, multiple sclerosis, epilepsy, hydrocephalus, enlarged ventricles, pathological lesions) or due to technical reasons (e.g. severe movement |
| | | | artifacts or inhomogeneity of the magnetic field). |
| **SHIP_TREND-0** | M-CIDI interview | Population based longitudinal cohort study | MDD subjects: no special exclusion criteria Control subjects: no lifetime diagnosis of depression, no antidepressiva, and severity index=0 All subjects: We removed subjects with due to medical conditions (e.g. a history of cerebral tumor, stroke, Parkinson's diseases, multiple sclerosis, epilepsy, hydrocephalus, enlarged ventricles, pathological lesions) or due to technical reasons (e.g. severe movement artifacts or inhomogeneity of the magnetic field). |
| **Singapore** | SCID interview | Inclusion: 1) DSM IV dx of MDD (Patients) 2) Age: 21-65 3) English speaking 4) Provision of informed written consent | Exclusion criteria 1) History of significant head injury 2)Neurological diseases such as epilepsy, cerebrovascular accident 3) Impaired thyroid function 4) Steroid use 5) DSM IV alcohol or substance use or dependence 6) Contraindications to MRI (e.g. pacemaker, orbital foreign body, recent surgery/procedure with metallic devices/implants deployed) using standard MRI Request Form from NNI 7)Pregnant women 8) Claustrophobia |
| **SoCAT** | SCID interview | Inclusion criteria: DSM IV dx for mdd patients Age: 18-65 right-handed currently depressed or remitted; Control subjects: any history of psychiatric disorder | Exclusion criteria 1) History of significant head injury 2)Neurological diseases such as epilepsy, cerebrovascular accident 3)Other diagnoses on Axis I disorders4) |
| **Stanford FAA** | SCID interview | Community-based DSM-diagnosed sample | MDD subjects: presence of axis-I disorders other than MDD, anxiety and eating disorders . Control subjects: control individuals did not meet diagnostic criteria for any current psychiatric. Both groups: alcohol / substance abuse or dependence within six months prior to MRI scanning, history of head trauma with loss of consciousness > 5 min, aneurysm, or any neurological or metabolic disorders that require ongoing medication or that may affect the central nervous system (including thyroid disease, diabetes, epilepsy or other seizures, or multiple sclerosis), MRI contraindications, or bad MRI data (e.g., extreme movement). |
| **Stanford T1w Aggregate** | SCID interview | Community-based DSM-diagnosed sample | MDD subjects: presence of axis-I disorders other than MDD, anxiety and eating disorders . Control subjects: control individuals did not meet diagnostic criteria for any current psychiatric. Both groups: alcohol / substance abuse or dependence within six months prior to MRI scanning, history of head trauma with loss of consciousness > 5 min, aneurysm, or any neurological or metabolic disorders that require ongoing medication or that may affect the central nervous system (including thyroid disease, diabetes, epilepsy or other seizures, or multiple sclerosis), MRI contraindications, or bad MRI data (e.g., extreme movement). |
| **TAD** | | | |
| **TIGER** | KSADS | Community-based DSM-diagnosed sample | All subjects: Exclusion criteria were premenarchal status (for females), history of concussion within the past 6 weeks or history |
| | | | of any lifetime concussion with loss of consciousness, contraindications to MRI scanning (e.g. braces, metal implants, or claustrophobia), serious neurological or intellectual disorders that could interfere with the participant's ability to complete study components. MDD subjects: meeting lifetime or current DSM-IV criteria for any Bipolar Disorder, Psychosis, or Alcohol Dependence, or DSM-5 criteria for Moderate Substance Use Disorder with substance-specific threshold for withdrawal. CTL subjects: any current or past DSM-IV Axis I Disorder and first-degree relative with confirmed or suspected history of depression, mania, psychosis, or substance dependence. |

**Supplementary Table 2**: ENIGMA MDD Image acquisition and processing by cohort

| Cohort | Scanner type | Sequence T1 | FreeSurfer version | Slice orientation | Operating system |
|---|---|---|---|---|---|
| **AFFDIS** | 3T Siemens Magnetom TrioTim | 3D T1 (176 slices; TR = 2250 ms; TE = 3.26 ms; FOV 256; voxel size 1X1X1mm) | 5,3 | Sagittal | Linux CentOS |
| **Barcelona** | 3T Philips Achieva | 3D MPRAGE images (Whole-brain T1-weighted); TR=6.7ms, TE=3.2ms; 170 slices, voxel size 0.89X0.89X1.2 mm. Image dimensions 288X288X170; field of view: 256X256X204; slice thickness: 1.2 mm; with a sagittal slice orientation, T1 contrast enhancement, flip angle: 8º, grey matter as a reference tissue, ACQ matrix MXP = 256X240 and turbo-field echo shots (TFE) = 218. | 6 | Sagittal | Scientific Linux 5 |
| **Cardiff** | A 3 Tesla whole body MRI system (General Electric, Milwaukee, USA) with an 8-channel head coil was used at the Cardiff University Brain Research Imaging Centre (CUBRIC). | High-resolution anatomical scan (Fast Spoiled Gradient-Recalled-Echo [FSPGR] sequence): 178 slices, TE=3 ms, TR=7.9 ms, voxel size=1.0×1.0×1.0 mm3, FA=15°, FOV=256×256 | 5,3 | | freesurfer-Linux-centos6_x86_64-stable-pub-v5.3.0 |
| **CSAN (Adf)** | 3T Siemens MAGNETOM PRISMA | Whole-head t1-weighted MPRAGE (TR = 2300 ms, TE = 2.34 ms, FOV 250 × 250 mm, voxel size = 0.9 × 0.868 × 0.868 mm, flip angle = 8°) | 7.2 | Sagittal | Ubuntu |
| **Calgary** | 1.5T Siemens Magnetom Vision. 3T GE Discovery MR750 | 1.5T: A sagittal scout series was acquired to test image quality. 3D fast low angle shot (FLASH) sequence was used to acquire data from 124 1.5 mm-thick contiguous coronal slices through the entire brain (echo time = 5ms, repetition time = 25ms, acquisition matrix = 256 x 256 pixels, field of view = 24 cm and flip angle = 40°). 3T: Anatomical imaging acquisition parameters: axial acquisition, repetition time (TR), 2200 milliseconds (ms); echo time (TE), 3.04 ms; TI, 766, 780; flip angle, 13 degrees; 208 partitions; 256 × 256 matrix; and field of view, 256. | 5,3 | Dalhousie sample, coronal; Calgary sample, axial | MacOs Sierra |
| **DCHS** | 3T Siemens Skyra | 3D multi-echo MPRAGE, voxel size 1 mm x 1mm x 1.5mm, TR = 2530 ms, TE = 1.69 x 3.55 x 5.41 x 7.27ms, FOV: 256x256mm, flip angle = 7° | 5.3 | Sagittal | Linux-centos6_x86_64 |
| **FOR2107 - Marbourg** | 3T Siemens Magnetom TiroTim syngo MR B17 | Sequence: 3D T1-weighted magnetization prepared rapid acquisition gradient echo (MPRAGE) - Sagittal Acquisition Direction, # of Slices 176, 0.5mm Slice Gap, 1.0x1.0x1.0 Voxel Size (mm3), TI 900 ms, TE 2.26 ms, TR 1900 ms, Flip Angle 9. | 5,3 | Sagittal | Red Hat Enterprise Linux Server release 5.11 (Tikanga) |
| **FOR2017 - Münster** | 3T Philips | 3D T1-weighted scan (170 slices; TR = 9ms; TE = 3.6ms; 256x231 matrix of 1×1×1 mm voxels) | 5,3 | Sagittal | Red Hat Enterprise Linux Server release 5.11 (Tikanga) |
| **FIDMAG** | 1.5T, GE Signa | 3D T1: matrix size = 512 × 512, 180 contiguous axial slices, voxel resolution = 0.47 × 0.47 × 1mm, no slice gap, TE = 3.93ms, TR = 2000ms and inversion time (TI) = 710ms, flip angle = 15 degrees | 6 | Axial | Linux-centos6_x86_64 |
| **Houston** | subjects in 20000s: 1.5 T Philips Medical Systems Gyroscan Intera; subjects in 30000s: 3T Siemens Allegra | Subjects in the 20000s: Fast field echo sequence- repetition time (TR) = 24 ms, echo time (TE) = 4.99 ms, flip angle = 40°, slice thickness = 1 mm, matrix size = 256 × 256 and 150 slices. Subjects in 30000s: MPRAGE- repetition time (TR) = 1750 ms, echo time (TE) = 4.39 ms, flip angle = 8°, slice thickness = 1 mm, matrix size = 208 × 256 and 160 slices. | 5,3 | Subjects in 20000s: Sagittal; Subjects in 30000s: Transverse | Fedora 19 |
| **Hiroshima** | 3T Siemens (Spectra, Verio.Dot), 3T GE (Signa HDxt) Site 1 = | T1 256x256x256 matrix of 1x1x1mm voxels (Siemens: ADNI MPRAGE (tfl), GRAPPA, 192 slices, GE: SPGR, 184 slices) *Detailed scanning parameter | 5,3 | Sagittal | Linux_Ubuntu_18.04 |
| | GE Signa HDxt 3.0T<br>2= GE Signa HDxt 3.0T<br>3 = SIEMENS MAGNETOM Spectra 3.0T<br>4 = SIEMENS MAGNETOM Verio.Dot 3.0T | sheets are available for all 4 scanners on request. | | | |
| **TiPs (Jena, Germany)** | 3T Siemens MAGNETOM Prisma_fit | MPRAGE sequence: TR 2300 ms, TE 3.03 ms, α 9°, 192 contiguous sagittal slices, in-plane field of view 256 mm, voxel resolution 1Å~1Å~1 mm; acquisition time 5:21 min | 5,3 | Sagittal | Linux |
| **MODECT** | 3T (General Electric Signa HDxt, Milwaukee, WI, USA) | T1-weigthed dataset was acquired (flip angle=12°, repetition time=7.84 milliseconds, echo time=3.02 milliseconds; matrix 256x256, voxel size 0.94x0.94x1 mm; 180 slices). | 5,3 | Coronal | Linux |
| **Melbourne** | 3T GE Signa Excite | 3D BRAVO sequence 140; TR=7900 ms; TE=3000 ms; flip angle=13º; FOV=256 mm; matrix=256 x 256 | 5,3 | Axial | Linux Debian x86 64 |
| **Minnesota** | 3.0 Tesla Tim Trio scanner; Siemens Corp | A 5-minute structural scan was acquired using a T1-weighted, high-resolution, magnetization-prepared gradient-echo sequence: repetition time, 2530 milliseconds; echo time, 3.65 milliseconds; inversion time, 1100 milliseconds; flip angle, 7°; field of view, 256 × 176 mm; voxel size, 1-mm isotropic; 224 slices; and generalized, autocalibrating, partially parallel acquisition acceleration factor, 2. | 5,3 | Coronal | Linux |
| **MOODS / DEP-ARREST CLIN** | 3T Philips Achieva | 3D T1-weighted image: TR=7, TE=3.5, FOV=352x352x180, Flip angle=8 degrees, number of slices : 180 slices, Slice gap 1 mm, voxel size: 0.8x0.8x1 | 6 | Transverse (Axial) | CentOS Linux 7 |
| **Moral Dilemma** | 3T GE Signa Excite | 3D BRAVO sequence: 140 contiguous slices; repetition time, 7900 ms; echo time, 3000 ms; flip angle, 13°; in a 25.6-cm field of view, with a 256 × 256 pixel matrix and a slice thickness of 1 mm (1 mm gap). | 5,3 | Axial | Linux Debian x86 64 |
| **Munster** | 3T Philips Gyroscan Intera | 3D fast gradient echo sequence (turbo field echo), repetition time = 7.4 milliseconds, echo time = 3.4 milliseconds, flip angle = 9°, two signal averages, inversion prepulse every 814.5 milliseconds, acquired over a field of view of 256 (feet -head [FH]) × 204 (anterior -posterior [AP]) × 160 (right -left [RL]) mm, phase encoding in AP and RL direction, reconstructed to | 5,3 | Sagittal | Red Hat Enterprise Linux Server release 5.11 (Tikanga) |

| | | cubic voxels of .5 mm × .5 mm × .5 mm | | | |
|---|---|---|---|---|---|
| **NESDA** | 3T Phillips Achieva/Intera | 3D gradient-echo T1-weighted sequence. TR=9 msec; TE=3.5 msec; flip angle 8º, FOV = 256 mm; matrix: 25x62x56; in plane voxel size = 1 mm × 1 mm x 1 mm; 170 slices. | 5 | Sagittal | SHARK HPC, Linux environment |
| **QTIM** | Bruker 4T Wholebody MRI | 3D T1 weighted sequence. TR=1500 msec; TE=3.35 msec; flip angle=8°, 256 or 240 (coronal or sagittal) slices, FOV=240 mm, matrix 256x256x256 (or 256x256x240) | 5,1 | Coronal, then sagittal following software upgrade. | Linux-centos4_x86_64-stable-pub-v5.1.0 |
| **San Francisco UCSF** | 3T GE Discovery MR750 | SPGR T1-weighted: TR=8.1 ms; TE=3.17 ms; TI=450 ms; flip angle=12°; 256x256 matrix; FOV=250x250 mm; 168 sagittal slices; slice thickness=1 mm; in-plane resolution=0.98x 0.98 mm | 5,3 | Sagittal | Linux-centos6_x86_64-stable-pub-v5.3.0. |
| **SHIP _START-2** | 1.5T Siemens Avanto | 3D T1-weighted (MP-RAGE/ axial plane); TR=1900 msec; TE=3.4 msec; Flip angle=15°; voxel size 1 mm x 1 mm x 1 mm | 5.3 (cortical), 5.1 (subcortical) | Axial | Centos6_x86_64 |
| **SHIP- _TREND-0** | 1.5T Siemens Avanto | 3D T1-weighted (MP-RAGE/ axial plane); TR=1900 msec; TE=3.4 msec; Flip angle=15°; voxel size 1 mm x 1 mm x 1 mm | 5.3 (cortical), 5.1 (subcortical) | Axial | Centos6_x86_64 |
| **Singapore** | Achieva 3T, Philips Medical Systems, Netherlands | Whole brain high resolution 3D MP-RAGE (magnetisation-prepared rapid acquisition with a gradient echo) volumetric scans (TR/TE/TI/flip angle 8.4/3.8/3000/8; matrix 256x204; FOV 240mm2) with axial orientation (reformatted to coronal) | 5,3 | Axial | Linux_Ubuntu12.04_6 4 |
| **SoCAT** | 3.0 T, Siemens Verio,Numaris/4,Syngo MR B17,Erlangen ,Germany | 3D T1 weighted MP-Rage/axial plane; TR=1900 msec; TE=3.4 msec; Flip angle=15°; Voxel size 1 mm x 1 mm x 1 mm | 5,3 | Axial | Ubuntu 18.04 LTS |
| **Stanford FAA** | 3.0T GE Discovery MR750 | Whole-brain T1-weighted images were collected using a spoiled gradient echo (SPGR) pulse sequence (186 sagittal slices; resolution = 0.9 mm isotropic; flip angle = 12°; repetition time [TR] = 6,240 ms; echo time [TE] = 2.34 ms) | 5,3 | Sagittal | Linux-centos6_x86_64 |
| **Stanford T1w Aggregate** | 1.5T GE Signa Excite | Whole-brain T1-weighted images were collected using a spoiled gradient echo (SPGR) pulse sequence (116 sagittal slices; through-plane resolution = 1.5 mm; in-plane resolution = 0.86 x 0.86 mm; flip angle = 15 degrees; repetition time [TR] = 8.3-10.1 ms; echo time [TE] = 1.7-3.0; inversion time [TI] = 300 ms; matrix = 256 x 192). | 5,3 | Sagittal | Centos6_x86_64, Linux-based HPC |
| **TAD** | | | | | |

| TIGER | 3T GE MR750 | TR/TE/TI=8.2/3.2/600 ms; flip angle=12°; 156 axial slices; FOV=25.6 cm; matrix=256 mm x 256 mm, isotropic voxel=1 mm, total scan time: 3:40 | 6 | Axial | Linux |
|---|---|---|---|---|---|

**Supplementary Table 3: List of hyperparameters of trained algorithms.** Optimal hyperparameters were found by the grid search during the sex classification task. We followed a heuristic approach outlined in (Hsu et al., 2003) to determine a range of values for C.

| Classification algorithm | Feature Selection | Hyperparameters | Nested CV |
|---|---|---|---|
| **SVM Linear** | None | C = $[10^{-4}, 10^{-3}, \ldots, 10^{4}]$ | 10 fold |
| **DenseNet** | None | Number of dense layers = [1,2,3]<br><br>Number of nodes in the dense layers = [10,100,200]<br><br>Adam optimizer: learning rate [.01,.001,.0001]<br><br>DropOut layer before dense layers (yes, no) | 10 fold |

**Supplementary Table 4: Comparison of SVM and DenseNet classification performance on entire dataset using integrated whole brain feature modalities.** The performance is evaluated via balanced accuracy, sensitivity, specificity, and AUC for each splitting strategy, with and without ComBat harmonization.

| | Splitting by Age/Sex | | Splitting by Site | |
|---|---|---|---|---|
| | No ComBat | With ComBat | No ComBat | With ComBat |
| **SVM** | | | | |
| Balanced Acc | 0.551 ± 0.021 | 0.478 ± 0.019 | 0.528 ± 0.039 | 0.520 ± 0.019 |
| Sensitivity | 0.477 ± 0.036 | 0.420 ± 0.024 | 0.490 ± 0.114 | 0.465 ± 0.033 |
| Specificity | 0.625 ± 0.030 | 0.536 ± 0.021 | 0.566 ± 0.124 | 0.574 ± 0.049 |
| AUC | 0.566 ± 0.021 | 0.490 ± 0.020 | 0.536 ± 0.062 | 0.520 ± 0.022 |
| **DenseNet** | | | | |
| Balanced Acc | 0.578 ± 0.022 | 0.561 ± 0.015 | 0.512 ± 0.019 | 0.508 ± 0.020 |
| Sensitivity | 0.452 ± 0.102 | 0.401 ± 0.090 | 0.428 ± 0.172 | 0.466 ± 0.265 |
| Specificity | 0.704 ± 0.104 | 0.721 ± 0.092 | 0.596 ± 0.217 | 0.550 ± 0.241 |
| AUC | 0.606 ± 0.026 | 0.595 ± 0.020 | 0.549 ± 0.076 | 0.544 ± 0.092 |

**Supplementary Table 5: MDD vs HC classification separated by hemispheres.** We evaluated whether a particular hemisphere could provide better classification accuracy.

| | Splitting by Age/Sex | | Splitting by Site | |
|---|---|---|---|---|

| Hemisphere | Left | Right | Left | Right |
|---|---|---|---|---|
| **SVM** | 0.546 ± 0.022 | 0.539 ± 0.017 | 0.514 ± 0.034 | 0.522 ± 0.036 |
| **DenseNet** | 0.569 ± 0.019 | 0.556 ± 0.024 | 0.513 ± 0.018 | 0.506 ± 0.017 |

**Supplementary Table 6: MDD vs HC classification stratified by main demographic and clinical characteristics.** Stratifications were applied to evaluate post-hoc whether groups that are more homogeneous would provide better classification accuracy.

| | | **Splitting by Site** | |
|---|---|---|---|
| | | SVM | DenseNet |
| **Sex** | Women (MDD n = 1,803; HC n = 2,383) | 0.516± 0.025 | 0.524± 0.018 |
| | Men (MDD n = 950; HC n = 1,857) | 0.513± 0.032 | 0.517± 0.044 |
| **Age of onset** | Adolescent (MDD n = 1,096; HC n = 4,240) | 0.547± 0.059 | 0.530± 0.102 |
| | Adults (MDD n = 1,302; HC n = 4,231) | 0.506± 0.070 | 0.579± 0.187 |
| **Number of episodes** | Recurrent (MDD n = 1,624; HC n = 4,240) | 0.543± 0.040 | 0.515± 0.043 |
| | Single (MDD n = 900; HC n = 4,240) | 0.492± 0.024 | 0.500± 0.020 |
| **AD use** | No (MDD n = 1,224; HC n = 4,240) | 0.503± 0.026 | 0.490± 0.037 |
| | Yes (MDD n = 1,313; HC n = 4,240) | 0.552± 0.049 | 0.544± 0.054 |

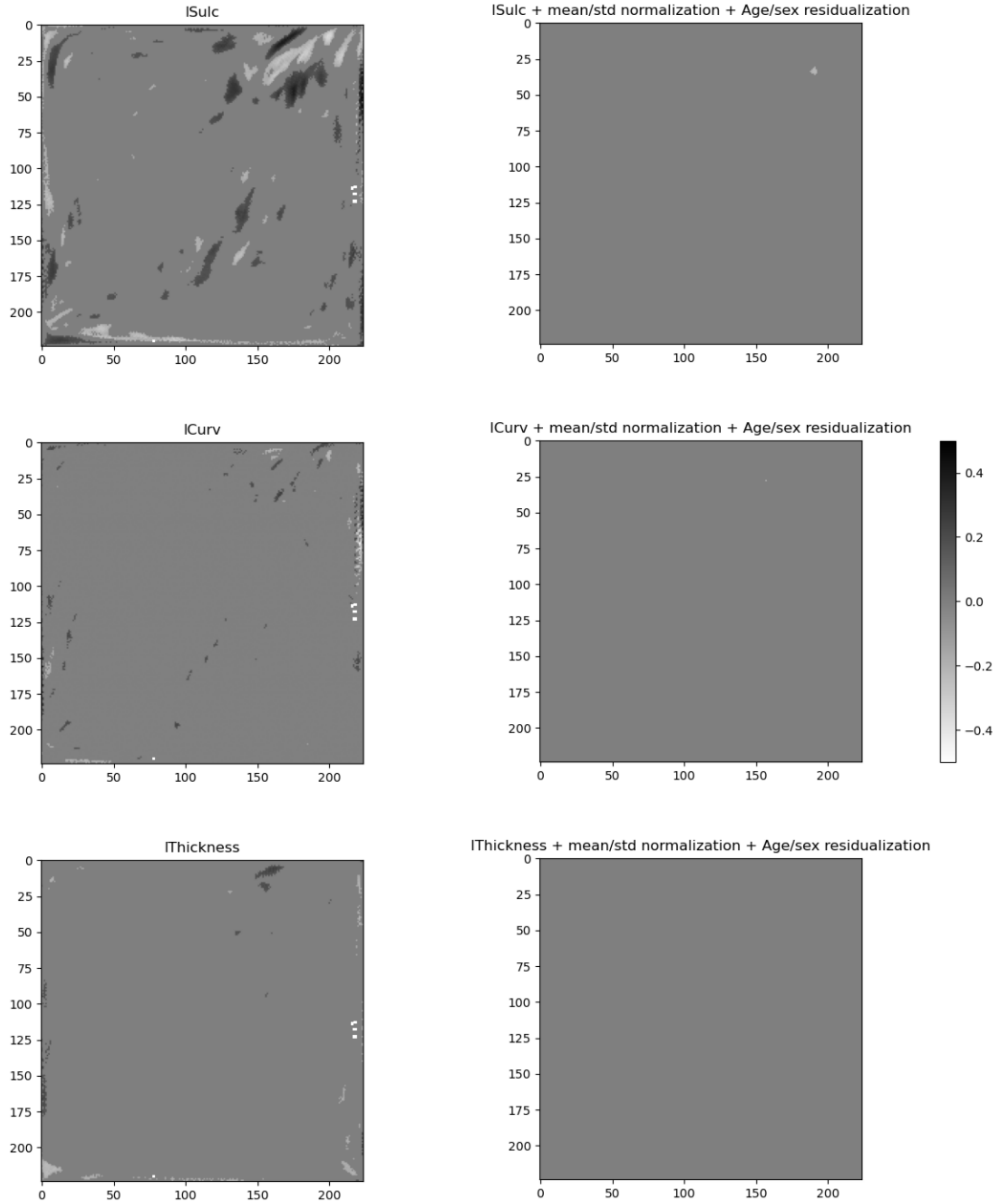

**Supplementary Figure 1: ICV dependence stored in the cortical maps.** To investigate if ICV is preserved in cortical features, we correlated ICV values with the pixels' intensities from SHIP_TREND_0 healthy controls (left). After standardizing the features to the mean of 0 and standard deviation of 1, and regressing out Age/Sex covariates, we effectively removed the effect of ICV from the features (right). Colormap represent the direction of the significant correlations.

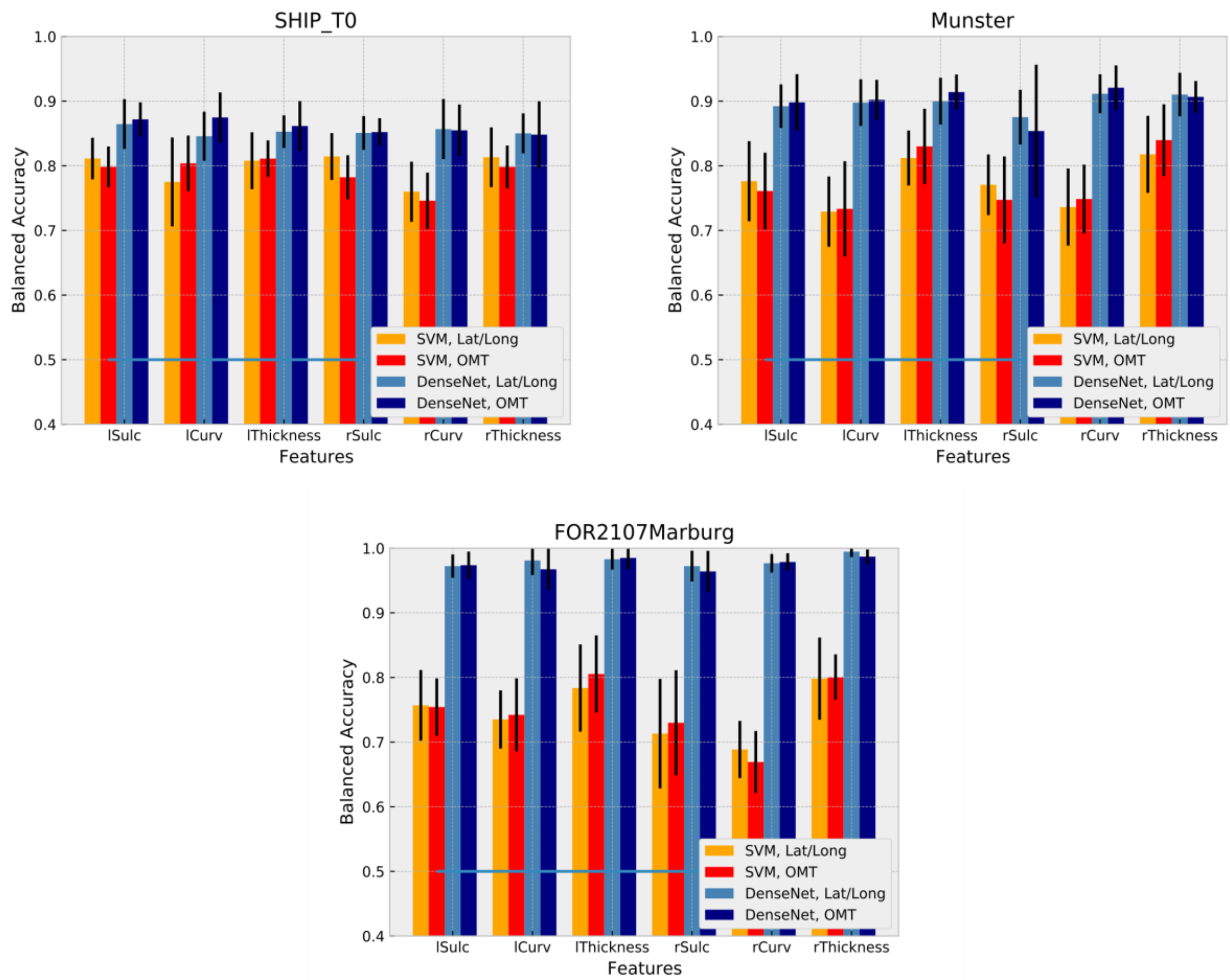

**Supplementary Figure 2: Sex classification.** We estimated classification performance via balanced accuracy of SVM and DenseNet on three biggest cohorts: SHIP_TREND-0 (top left), Munster (top right) and FOR2107Marburg (bottom) for all features separately using 1) Latitude/Longitude projection and 2) OMT projection.

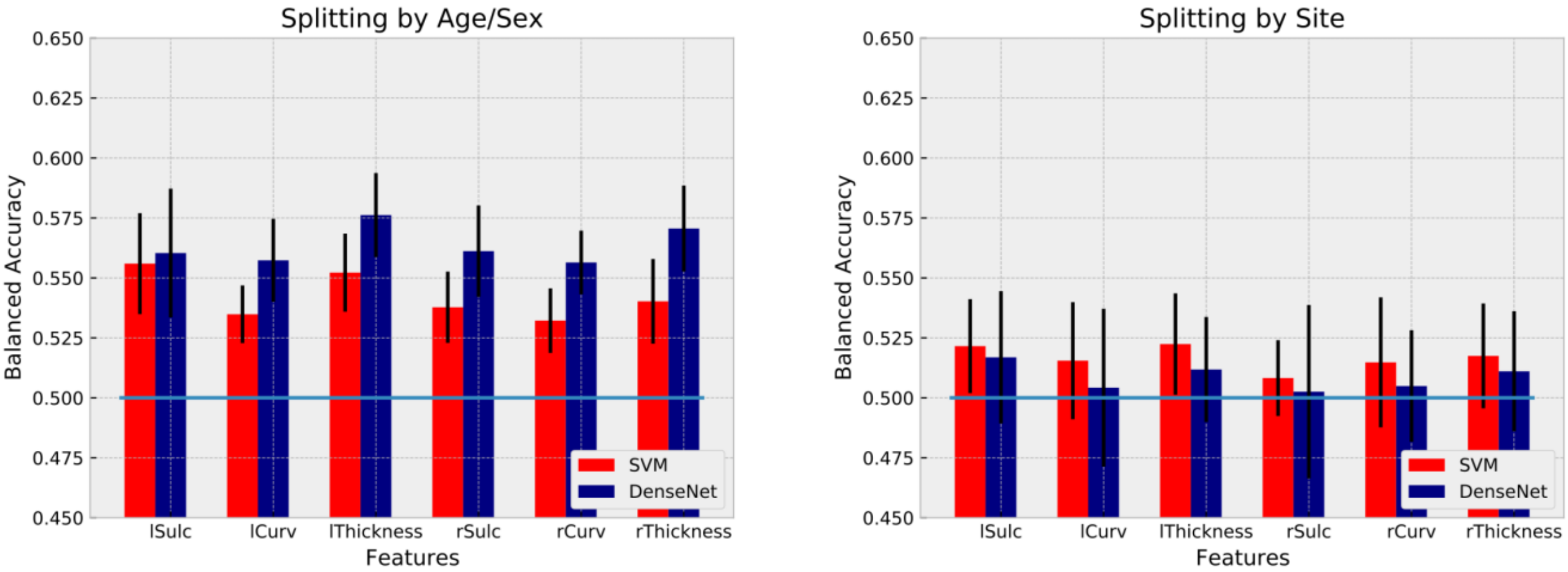

**Supplementary Figure 3: MDD vs HC classification.** We estimated classification performance via balanced accuracy of SVM and DenseNet for each hemisphere and feature type separately.

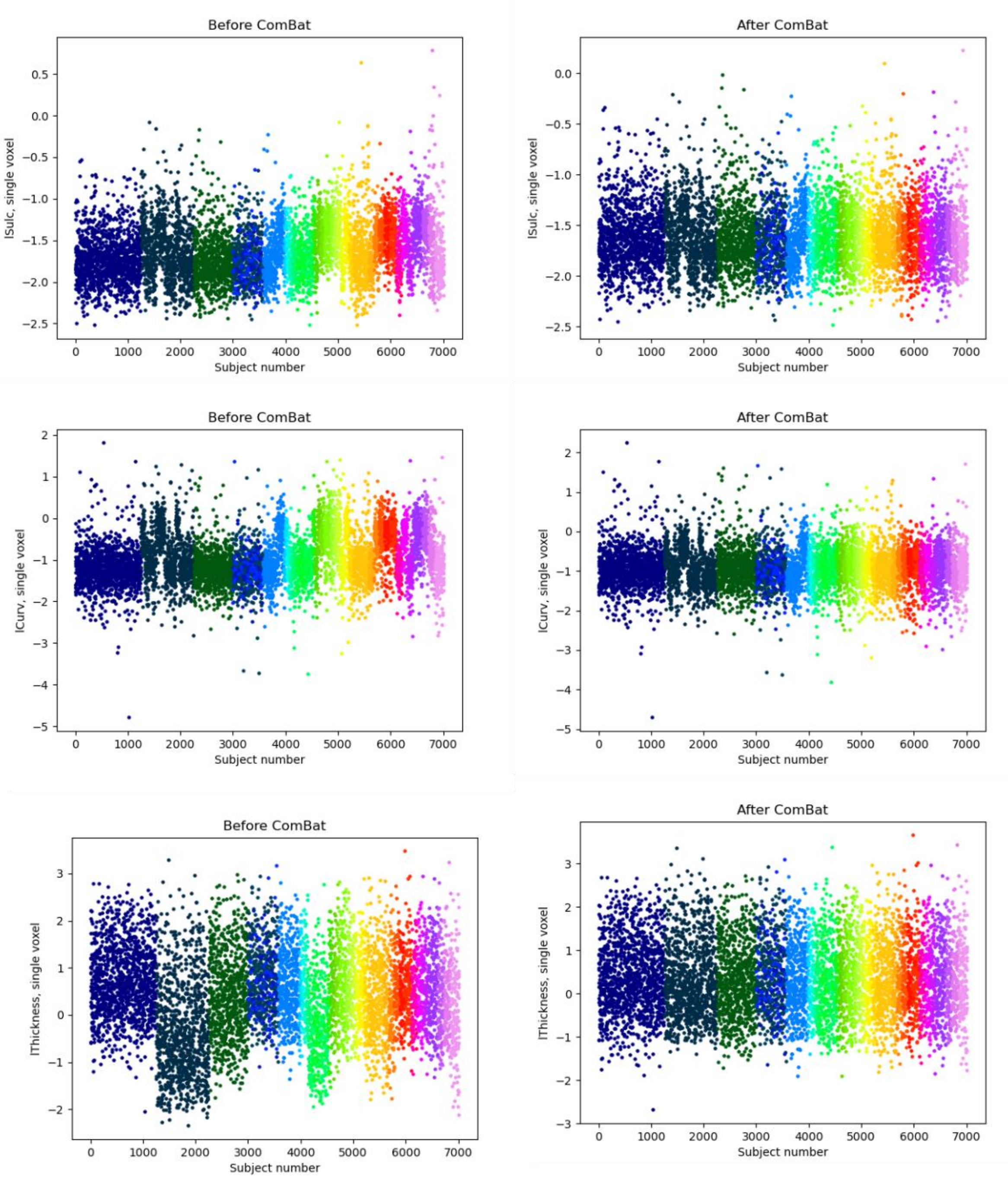

**Supplementary Figure 4: Examples of ComBat harmonization for all data modalities.** Color corresponds to the site affiliation.